\documentclass[
	aps,prd,
	reprint,
	showkeys,
	amsmath,amssymb,
	nofootinbib,
	floatfix
	]{revtex4-2}
\usepackage[utf8]{inputenc}
\usepackage[T2A]{fontenc}
\usepackage{aas_macros}

\usepackage{hyperref}
\hypersetup{ %
			colorlinks=true,%
			pdfauthor={Michael Rashkovetskyi, Julian B.~Mu\~noz, Daniel J.~Eisenstein, and Cora Dvorkin},%
			pdftitle={Small-scale clumping at recombination and the Hubble tension},%
			citecolor=blue%
			}
\usepackage{graphicx}
\usepackage{xcolor}

\begin{document}

\title{Small-scale clumping at recombination and the Hubble tension}
\author{Michael Rashkovetskyi}
\email{mrashkovetskyi@cfa.harvard.edu}
\affiliation{Center for Astrophysics | Harvard \& Smithsonian, 60 Garden St, Cambridge, Massachusetts 02138}
\author{Julian B.~Mu\~noz}
\affiliation{Center for Astrophysics | Harvard \& Smithsonian, 60 Garden St, Cambridge, Massachusetts 02138}
\author{Daniel J.~Eisenstein}
\affiliation{Center for Astrophysics | Harvard \& Smithsonian, 60 Garden St, Cambridge, Massachusetts 02138}
\author{Cora Dvorkin}
\affiliation{Department of Physics, Harvard University, 17 Oxford Street, Cambridge, Massachusetts 02138}
\date{\today}

\begin{abstract}
Despite the success of the standard $\Lambda$CDM model of cosmology, recent data improvements  have made tensions emerge between low- and high-redshift observables, most importantly in determinations of the Hubble constant $H_0$ and the (rescaled) clustering amplitude $S_8$.
The high-redshift data, from the cosmic microwave background (CMB), crucially relies on recombination physics for its interpretation.
Here we study how small-scale baryon inhomogeneities (i.e., clumping) can affect recombination and consider whether they can relieve both the $H_0$ and $S_8$ tensions.
Such small-scale clumping, which may be caused by primordial magnetic fields or baryon isocurvature below kpc scales, enhances the recombination rate even when averaged over larger scales, shifting recombination to earlier times.
We introduce a flexible clumping model, parametrized via three spatial zones with free densities and volume fractions, and use it to study the impact of clumping on CMB observables.
We find that increasing $H_0$ decreases both $\Omega_m$ and $S_8$, which alleviates the $S_8$ tension.
On the other hand, the shift in $\Omega_m$ is disfavored by the low-$z$ baryon-acoustic-oscillations measurements.
We find that the clumping parameters that can change the CMB sound horizon enough to explain the $H_0$ tension also alter the damping tail, so they are disfavored by current {\it Planck} 2018 data. 
We test how the CMB damping-tail information rules out changes to recombination by first removing $\ell>1000$ multipoles in {\it Planck} data, where we find that clumping could resolve the $H_0$ tension.
Furthermore, we make predictions for future CMB experiments, as their improved damping-tail precision can better constrain departures from standard recombination.
Both the {\it Simons Observatory} and CMB-S4 will provide decisive evidence for or against clumping as a resolution to the $H_0$ tension.
\end{abstract}


\maketitle

\section{Introduction}


The standard $\Lambda$-cold dark matter ($\Lambda$CDM) model of cosmology has proven to be remarkably successful in interpreting different measurements consistently and simultaneously.
These include the cosmic microwave background (CMB, for example \cite{WMAP,planck_cosmo}), the large-scale structure (LSS, e.g. \cite{BAO-discovery,BAO-constraints,DES-Y3}), and probes of the expansion rate of the Universe (such as \cite{accelerated-expansion,SH0ES}).
However, as the precision of these probes has increased, tensions between them have started to appear.

A notable problem that has emerged within $\Lambda$CDM is the Hubble tension---a discrepancy between the cosmic expansion rate today (given by the Hubble parameter $H_0$) inferred from different data sets.
On the one side, the standard CMB analysis of {\it Planck} 2018 data yields a value of $H_0=\left( 67.4\pm 0.5 \right)$ km s$^{-1}$ Mpc$^{-1}$ \citep{planck_cosmo}.
On the other side, direct $H_0$ measurements (from type Ia supernovae calibrated with Cepheids \citep{carnegie-hubble,carnegie-supernovae-Ia,near-IR-std-candles,SH0ES} or surface brightness fluctuations \citep{supernovae-Ia-SBF}; from type II supernovae \citep{supernovae-II}, strong-lensing time delays \citep{H0LiCOW,TDCOSMO,STRIDES}, gravitational waves standard sirens \citep{GW-std-sirens}, Tully-Fisher relations \citep{tully-fisher-rel}, tip of the red giant branch \citep{TRGB}, Mira variables \citep{Mira-var} or megamasers \citep{MCP}) give higher values.
In this paper, we will focus on the latest distance-ladder measurement of the {\it Supernovae, H0, for the Equation of State of Dark Energy} (SH0ES) Collaboration $H_0=\left( 73.2\pm 1.3 \right)$ km/(s Mpc) \citep{SH0ES}, which is $4.2\sigma$ away from {\it Planck}.

There is another, weaker tension in the values of the matter fraction $\Omega_m$ and the amplitude of galaxy clustering $\sigma_8$ (on spheres of comoving radius $R=8/h$ Mpc).
Instead of $\sigma_8$, a related parameter $S_8=\sigma_8\left(\Omega_m/0.3\right)^{0.5}$ is often used, as it is less correlated with $\Omega_m$ in LSS data.
\citet{planck_cosmo} report $\Omega_m=0.315\pm 0.007$ and $S_8=0.831\pm 0.017$ with {\it Planck} data, while the Dark Energy Survey Year 1 (DES-Y1, \cite{DES-Y1}) weak-lensing and galaxy-clustering data obtains $\Omega_m=0.264^{+0.032}_{-0.019}$, $S_8=0.783^{+0.021}_{-0.025}$.
The new DES Year 3 results are $\Omega_m=0.339^{+0.032}_{-0.031}$, $S_8=0.776\pm0.017$ \citep{DES-Y3}, making the tension weaker, but still worth exploring,
as other LSS probes disagree with the CMB~\citep{KiDS-1000,unWISE-plancklensing,dens-perturb-growth,cmb-lss-disagree1,cmb-lss-disagree2,cmb-lss-disagree3}.

Various extensions to $\Lambda$CDM have been proposed to solve the $H_0$ discrepancy. They can be broadly divided into early- and late-type solutions, with the former changing the length of the standard ruler \citep{early-sol19ede,early-sol19rock,early-sol19ade,early-sol20edenu,early-sol20nu}, and the latter the evolution of the expansion rate $H(z)$ at low redshifts \citep{late-sol17,late-sol18,late-sol20a,late-sol20b}. Only the early-type solutions can be in agreement with low-$z$ standard-ruler measurements of the BAOs, though the most popular model of early dark energy \citep{EDE-issue} worsens the $S_8$ tension (for a recent review see \citet{hubble-tension-guide}). 
Here, instead, we study how changing the physics of recombination is an early-type solution to both the $H_0$ and $S_8$ tensions, as first proposed in \cite{JP20}.

The interpretation of CMB data crucially relies on the physics of recombination, so it is natural to ask how well understood, and constrained, this transition is.
The process of hydrogen recombination depends crucially on the two-body recombination rate, and thus can be affected by physics at very small scales \citep{PMF11}.
In \cite{PMF04,PMF13,PMF19} it was shown that, by creating baryonic clumping at small scales, primordial magnetic fields (PMFs, for a review see \citet{PMF-review}) would leave an imprint on the CMB by allowing a more-rapid process of recombination, and shifting the decoupling between photons and baryons to larger $z$.
While such inhomogeneities would take place at very small scales, they enhance the recombination rate when averaged over larger scales. 
An earlier recombination implies a higher $H_0$ for a fixed angular sound horizon $\theta_s$ in the CMB. Recently, \citet{JP20} have shown that such a clumping could relieve both the Hubble and $S_8$ tensions in current cosmological data.
On the other hand, \citet{clumping-ACT} argued that the $H_0$ value inferred from {\it Planck} and Atacama Cosmology Telescope (ACT) data remains in significant tension with SH0ES.

Here we extend previous analyses by introducing a very generic model of clumping at small scales. This model posits that baryons live in three zones: an average one, and an over/underdense one (see Fig.~\ref{fig:3zone_model}). This encompasses the models in \citet{JP20} and \citet{clumping-ACT}, as well as other possible origins of small-scale baryonic clumps, such as baryon isocurvature \citep{baryon-isocurvature}.

The key question we tackle is whether a change in recombination that is sufficient to change the sound horizon---and thus explain the $H_0$ tension---leaves a detectable imprint on the CMB damping tail. 
The high-$\ell$ CMB tail has been measured to great success by the {\it Planck}, ACT \citep{ACT}, and South Pole Telescope (SPT, \cite{SPT}) collaborations; and upcoming measurements from the {\it Simons Observatory} (SO, \cite{SO}) and CMB-S4 \citep{CMBS4white} will improve those measurements even further.


This paper is structured as follows.
We start in Sec.~\ref{sec:M3model} by defining our generalized clumping model ({\bf M3}), and discussing its physical implications in Sec.~\ref{sec:effects}.
In Sec.~\ref{sec:planck} we analyze the current CMB data from {\it Planck} 2018 \citep{planck_ttee,planck_lensing}, exploring clumping-$H_0$ correlations as well as looking into shifts in $S_8$ and $\Omega_m$. 
Then we perform forecasts for future CMB experiments in Sec.~\ref{sec:forecasts}, to understand how better measurements of the damping tail will test our clumping model more precisely.
We conclude in Sec.~\ref{sec:conclusion}.
    
\section{Three-zone model (M3) for recombination}
\label{sec:M3model}

\begin{figure}[ht!]
\includegraphics[width=\columnwidth]{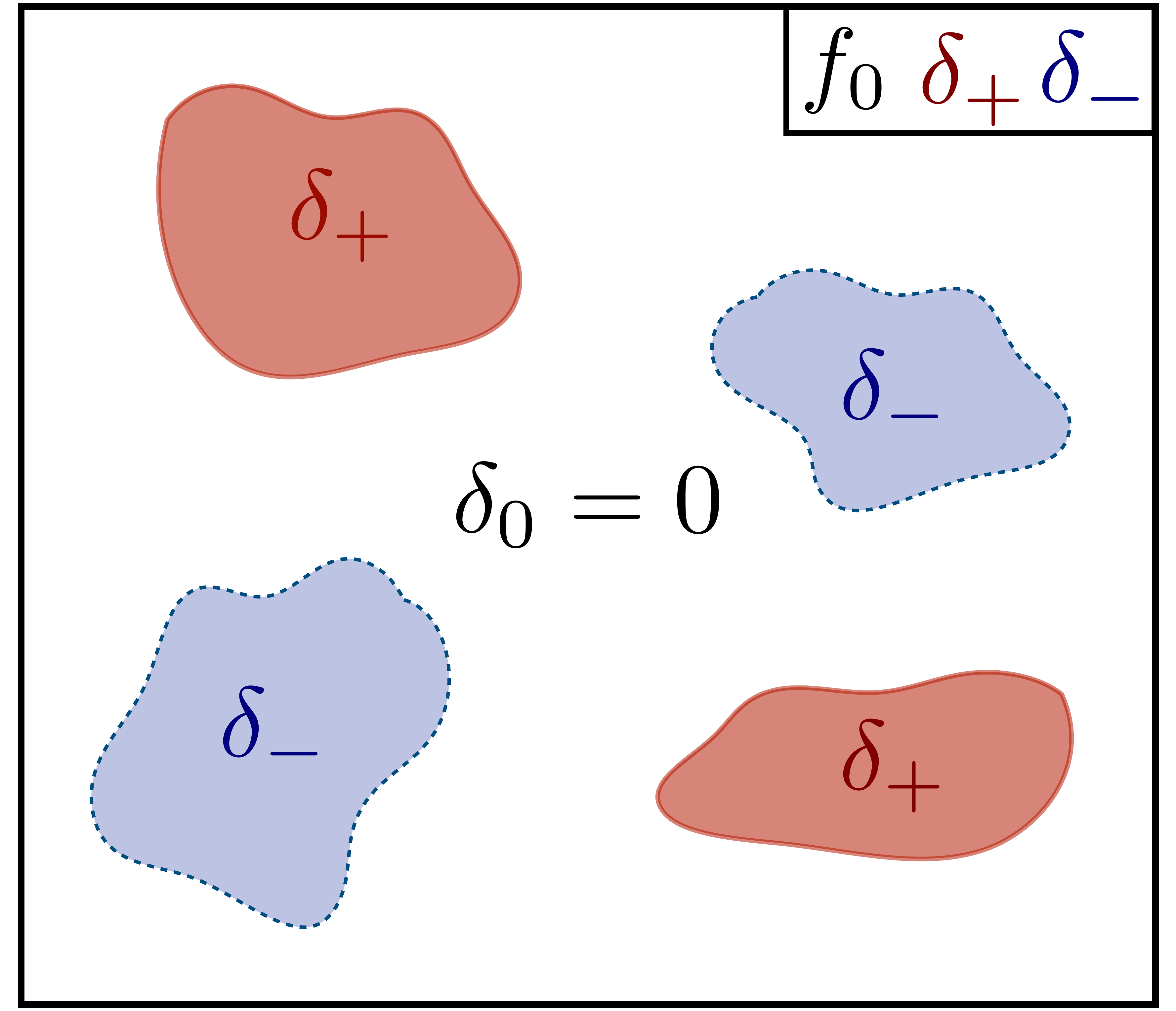}
\caption{M3 spatial structure: regions with average ($0$), lower ($-$) and higher ($+$) density.
We take those three effective zones to have constant overdensity $\delta_i=n_{H,i}/\left\langle n_H\right\rangle-1$ each, and the structure is fixed in time.
The model is described by six parameters: three $\delta_i$'s and three volume fractions $f_i$ that each zone occupies.
There are three constraints: $\sum_i f_i=1$, $\sum_i f_i \delta_i=0$ and $\delta_0=0$.
So we choose $f_0$ and $\delta_\pm$ as input parameters, as highlighted in upper right corner.
}
\label{fig:3zone_model}
\end{figure}

We begin by defining our three-zone model (M3) for baryonic clumping.
The general idea is that there are fluctuations on very small ($\sim$ kpc for PMFs~\citep{PMF-review}) scales, so that the large-scale behavior of baryons follows the usual assumptions; whereas the recombination rate, which depends on the electron density squared ($n_e^2$) can be enhanced (for large overdensities) or reduced (for underdensities), with respect to the average.

Our general picture is illustrated in Fig.~\ref{fig:3zone_model}: there are regions with average (marked by the index $0$), lower ($-$) and higher ($+$) density. 
For simplicity, we take those three effective zones to have constant hydrogen density $n_H$ each, and the structure is constant in time.
Then one needs six parameters: three weights (volume fractions) $f_i$ and three densities, which can be parametrized as $\Delta_i=n_{H,i}/ \left\langle n_H \right\rangle$ or $\delta_i=\Delta_i-1$ (hereafter $i=\left\{ 0,-,+ \right\}$).
However, we set three constraints: first, all volume is divided between the three zones ($\sum_i f_i=1$), second, the total baryonic mass is set by $\omega_b$ ($\sum_i f_i \Delta_i=1$, or equivalently $\sum_i f_i \delta_i=0$), and finally, one of the zones has average density ($\delta_0=0$, which is optional but simplifies the analysis).
This leaves three free parameters.
For input, we choose the two nonzero relative overdensities $\delta_-$, $\delta_+$ and the volume fraction $f_0$ of the average-density zone.

This three-parameter model is very flexible, as for example it encompasses the M1 and M2 models presented in \citet{JP20} (obtained by fixing $f_0=1/3$, and either $\delta_-=-0.9$ for M1 or $\delta_-=-0.7$ for M2).
However, our M3 model is only bound by the constraints of volume and mass conservation and the only arbitrary choice is setting one of the three regions to have average density.
On the flip side, this flexibility makes the three parameters very degenerate: if either $f_0\rightarrow 1$, $\delta_+$ or $\delta_-\rightarrow 0$, then the deviation from uniform density becomes negligible.

Therefore, the prior on these parameters ought to be balanced between generality and degeneracy. 
We choose a log-uniform prior on $\left| \delta_- \right|$ ($10^{-5}\le \left| \delta_- \right|\le 0.955$) and on the ratio $\left| \delta_+/\delta_- \right|$ ($0.1\le\left| \delta_+/\delta_- \right|\le 10$), and a uniform prior on $f_0$ ($0\le f_0\le 1$).
The lower bound on $|\delta_-|$ is set to the curvature perturbations $\sqrt{A_s}\sim 10^{-5}$, whereas the higher bound is determined by numerical limitations of the recombination code, more extreme underdensities cause the integration in RECFAST to fail.
Bounds on the ratio are chosen so that the under and overdensity regions are within an order of magnitude from each other, to avoid the unnatural configuration when one $\delta$ is negligible and other is significant.
Moreover, the constraints force the volume fractions ratio to be $f_+/f_-=-\delta_-/\delta_+$, so very different $\delta$'s will cause one of the volume fractions to be tiny and the whole clumping effect negligible.

An important parameter quantifying the amplitude of the inhomogeneities is the relative variance of densities,
\begin{equation}
b=\frac{\left\langle \left( n_H - \left\langle n_H \right\rangle \right)^2 \right\rangle}{\left\langle n_H \right\rangle^2} = -\delta_- \delta_+ \left( 1-f_0 \right),
\label{eq:b-def}
\end{equation}
hereafter denoted as the clumping parameter $b$, following \citet{JP20}.

Technically, we implement this model in a fork\footnote{\url{https://github.com/misharash/class\_public}} of the CLASS code\footnote{\url{https://lesgourg.github.io/class_public/class.html}} \citep{CLASS}. 
We run the standard recombination code RECFAST \citep{recfast1,recfast2,recfast3} within each zone separately, given its density $n_{H,i}$, producing three recombination histories, in terms of their free electron fractions $x_{e,i}\left( z \right)=n_{e,i}/n_{H,i}$.
These are then averaged 
\begin{equation}
x_e\left( z \right) = \sum_i f_i\Delta_i x_{e,i}\left( z \right)
\end{equation}
and passed to the rest of modules in CLASS.

We also have tested our method with the more precise recombination code HyREC~\citep{hyrec2011}. 
While the two codes differ in their predictions of the highest $\ell$ modes (see e.g.~\citet{hyrec2}), we find this difference subleading for the purposes of this paper, as shown in Appendix \ref{sec:justify-reccodes}.

\section{The effects of small-scale gas clumping on the CMB}
\label{sec:effects}

\begin{figure}[ht!]
\includegraphics[width=\columnwidth]{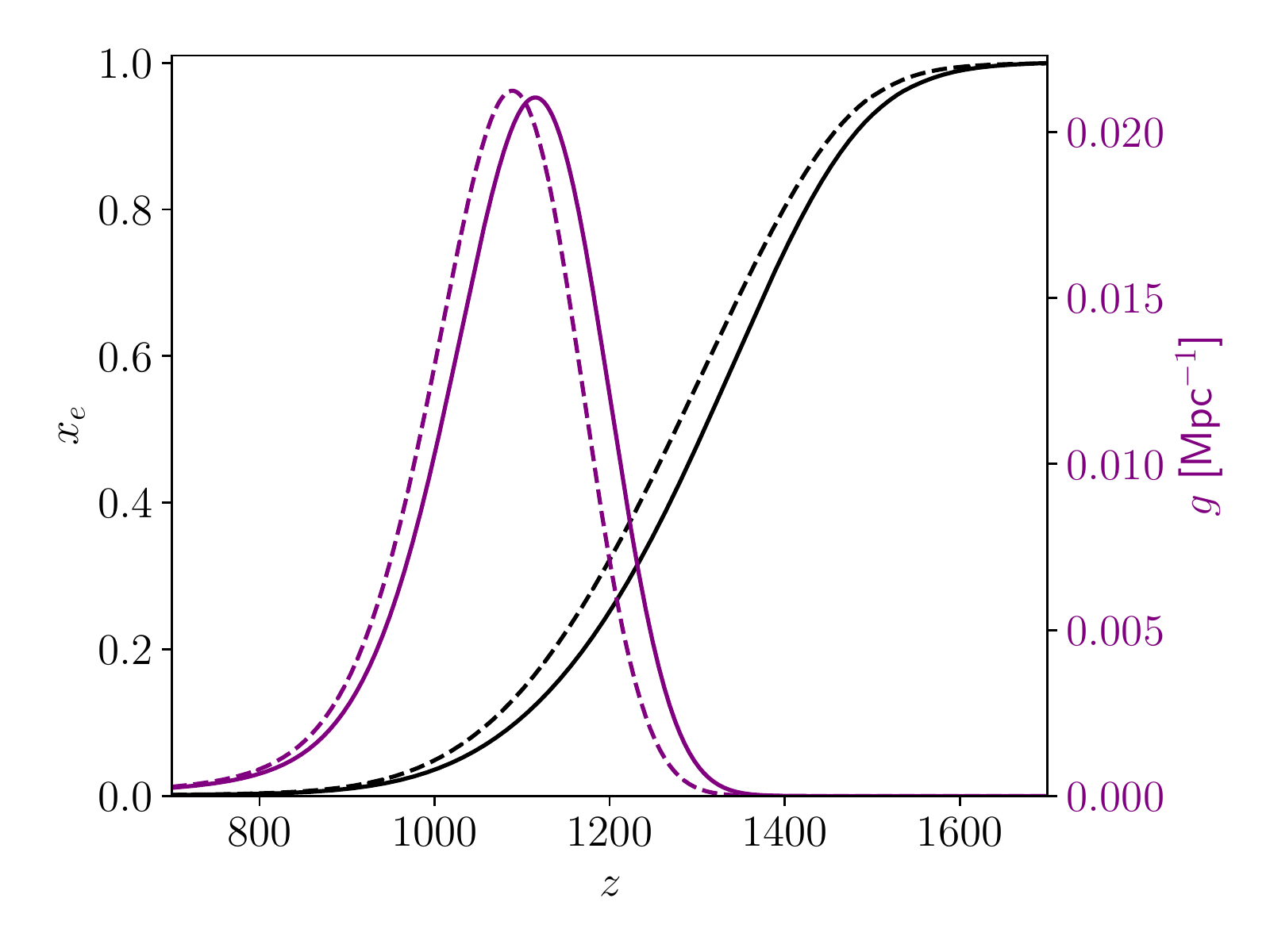}
\caption{Demonstration of the effect of clumping on the recombination history ($x_e$, black) and visibility function ($g$, purple) versus redshift $z$.
Standard recombination is shown as dashed curves, and clumping in solid, where we have set $\delta_-=-0.9$, $\delta_+=5/3$, and $f_0=1/3$, to yield $b=1$.
Cosmology ($\theta_s$, $\omega_b$, $\omega_\mathrm{cdm}$, $A_s$, $n_s$, $\tau_\mathrm{reio}$) is fixed to the {\it Planck} best fit \citep{planck_cosmo}.}
\label{fig:M3demo-rec}
\end{figure}

\subsection{Recombination}

Consider the recombination of an effective three-level hydrogen atom,
\begin{equation}
\frac{dn_e}{dt}+3Hn_e=-\left( \alpha n_en_{H^+}-\beta n_{H^0}e^{-E_{21}/kT} \right)C,
\label{eq:3level-rec}
\end{equation}
where $\alpha$ and $\beta$ are recombination and photoionization rate coefficients, $E_{21}$ is the energy difference between the first excited level and the ground state, $k$ is Boltzmann constant, $T$ is temperature and $C$ is an additional factor taking into account both Lyman-$\alpha$ and two-quantum decays \citep{reco-peebles,reco-zeldovich}.
If we take a spatial average, all terms except the first on the right-hand side, will depend on $\left\langle n_H \right\rangle$, while that one term will depend on $\left\langle n_H^2 \right\rangle$, as it corresponds to recombinations (binding of an electron and an ion). 
Introducing inhomogeneities enhances the average recombination rate, as $\left\langle n_H^2 \right\rangle \ge \left\langle n_H \right\rangle^2$ (where equality is only reached for uniform density). 
This causes $n_e$ to decrease faster, and the Universe to become neutral and transparent to radiation earlier than in the homogeneous case.
Given a recombination history, we define the visibility function
\begin{equation*}
g = \dot\tau e^{-\tau}
\end{equation*}
as the probability that a CMB photon last scattered per unit conformal time, and thus determines the effective redshift of recombination.
It is given in terms of the Thomson optical depth $\tau$ and its derivative with respect to conformal time $\eta$ (namely the inverse of photon's comoving mean free path),
\begin{equation}
\dot\tau = \sigma_T n_{H,\rm now} x_e \left( 1+z \right)^2, \label{eq:inv-mfp-comov}
\end{equation}
where $n_{H,\rm now}$ is hydrogen number density today, which is not time- or redshift-dependent, and $\sigma_T$ is Thomson scattering cross-section.

While Eq.~\eqref{eq:3level-rec} is a simple approximation---and in our implementation we include the detailed physics of the recombination codes---it serves to illustrate how small scales affect recombination at large scales. 
As an example, Fig.~\ref{fig:M3demo-rec} shows how clumping affects recombination, where the intuition from Eq.~\eqref{eq:3level-rec} remains true: clumping shifts recombination and thus the peak of the visibility function to higher redshifts.

\begin{figure}[ht!]
\includegraphics[width=\columnwidth]{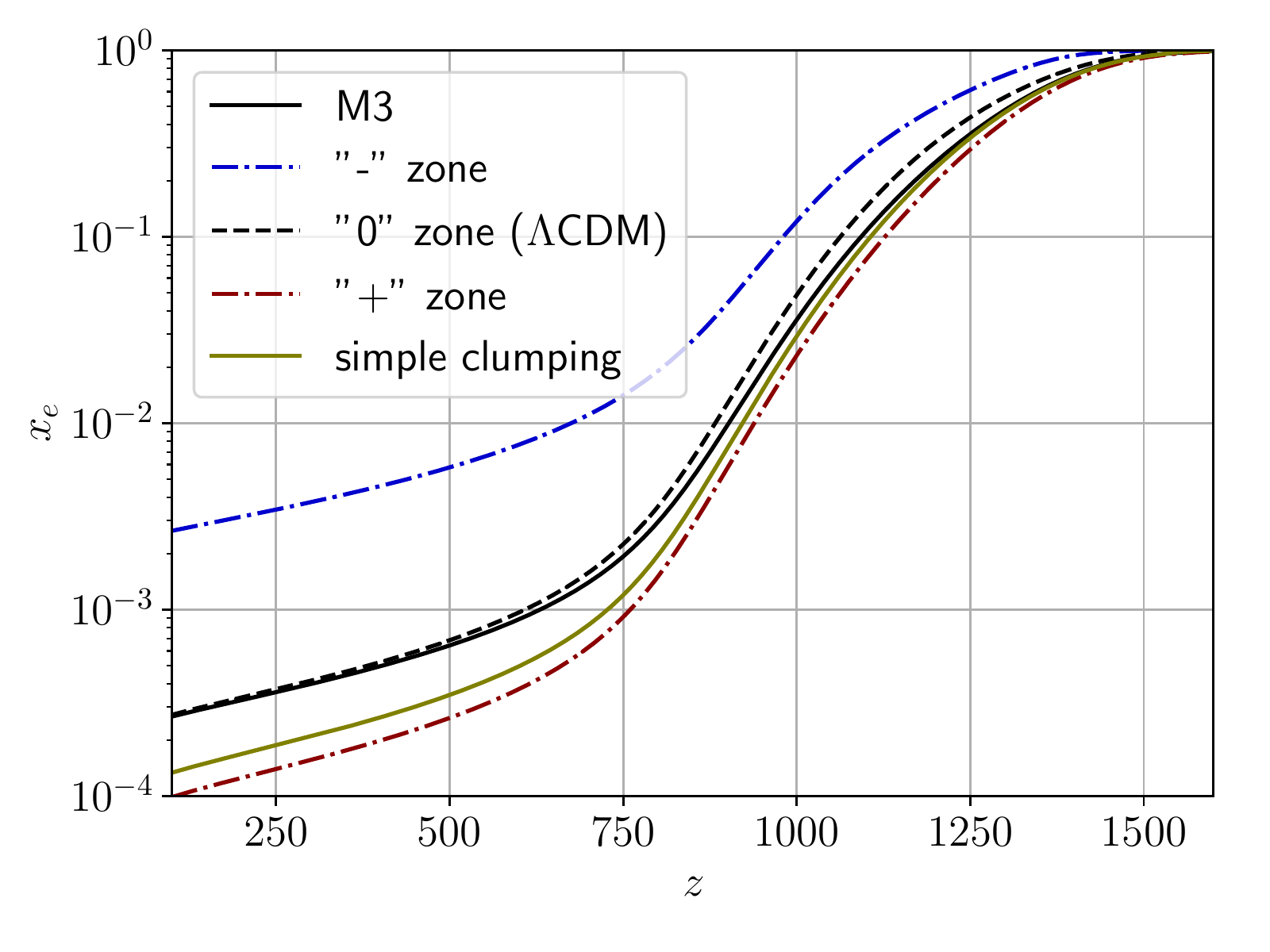}
\caption{Ionization fraction in different zones of the M3 and simple-clumping models.
The clumping parameters are $\delta_-=-0.9$, $\delta_+=5/3$, and $f_0=1/3$, $b=1$.
Cosmology ($\theta_s$, $\omega_b$, $\omega_\mathrm{cdm}$, $A_s$, $n_s$, $\tau_\mathrm{reio}$) is fixed to the {\it Planck} best fit \citep{planck_cosmo}.
The ``-'' zone behaves very different from the others and has a notable low-redshift tail.
The total M3 ionization fraction approaches the ``0'' zone (namely standard recombination), while for simple clumping (when rates of recombination and other two-body processes are just enhanced by a factor of $1+b$) $x_e$ stays lower.
The simple-clumping model produces larger deviations in power spectra than M3 for the same $H_0$ change and is only shown here for comparison purposes.}
\label{fig:M3demo-rec-details}
\end{figure}

To show how recombination evolves in over/under-dense regions, we plot the ionization fraction in each of the three zones of our M3 model in Fig.~\ref{fig:M3demo-rec-details}.
The underdense (``-'') zone has a dramatically delayed recombination history, and it presents a sizable low-$z$ tail.
The total M3 ionization fraction approaches the ``0'' zone (namely standard recombination) for lower redshifts.

We note that manually enhancing the average recombination rate (i.e., without keeping track of the overdense and underdense zones) does not capture the entire effect of clumping.
To illustrate that, we have also implemented a simple one-parameter clumping model, using only $b$ [Eq.~\eqref{eq:b-def}], where we have assumed a spatially uniform $x_e$ and have therefore just multiplied the rates of recombination and other two-body processes by $\left\langle n_H^2\right\rangle/\left\langle n_H\right\rangle^2=1+b$ inside RECFAST.
From Fig.~\ref{fig:M3demo-rec-details} it is clear that  this simple model behaves like an overdense zone and does not reproduce the behavior of M3.
Such a difference arises because the recombination rate is proportional to $\left\langle n_en_{H^+}\right\rangle\approx \left\langle x_e^2n_H^2\right\rangle$.
Only if one assumes constant $x_e$ can one simply put the latter equal to $\left\langle x_e\right\rangle^2\left\langle n_H^2\right\rangle=\left\langle x_e\right\rangle^2\left\langle n_H\right\rangle^2\left(1+b\right)$.
But if we assume no electron mixing between the zones, at each density the recombination goes at its own pace.
At lower densities the ionization fraction is higher and vice versa.
Then $\left\langle x_e^2n_H^2\right\rangle<\left\langle x_e\right\rangle^2\left\langle n_H\right\rangle^2\left(1+b\right)$, and the actual ratio is time-dependent.
As a consequence, this simple clumping model causes significantly higher difference in power spectra $C_\ell$ for the same change in $H_0$ than M3 and thus cannot better alleviate the Hubble tension.
This implies that matching the low-$z$ tail of recombination is key for consistency with CMB data.
Therefore only M3 and not simple clumping is used further in the paper.

\subsection{Sound horizon at last scattering}

Shifting the epoch of recombination affects the quantities derived from the CMB.
For example, distances are inferred from well-measured CMB angular scales, such as the angular scale of the sound horizon at last scattering $\theta_s = r_s/r_*$, where
\begin{equation}
\label{eq:ls-distance}
r_* = \int_0^{z_*} \frac{cdz}{H\left( z \right)}
\end{equation}
is the comoving distance to the last scattering surface, $c$ is speed of light, and
\begin{equation}
\label{eq:ls-soundhor}
r_S = \int_{z_*}^\infty \frac{c_S\left( z \right)dz}{H\left( z \right)}
\end{equation}
is the comoving distance a sound wave (of speed $c_S$) could travel before last scattering, called sound horizon. 
The redshift $z_*$ of last scattering is determined by recombination physics (in particular by the peak of the visibility function) and depends on the radiation, baryon, and matter densities,
\begin{equation*}
z_* = z_*\left( \omega_r,\omega_b,\omega_m \right); \, \omega_j=\Omega_jh^2,
\end{equation*}
where $h=H_0/[100\,{\rm km/(s/Mpc)}]$.

The angular sound horizon $\theta_s \sim 1/\ell_{\rm peak}$ is measured very well from the CMB power spectrum, where
$\ell_{\rm peak}$ is the multipole of the first acoustic peak.
This leaves two main avenues to obtain a larger $H_0$ from the CMB and solve the Hubble tension.
Late-type solutions change the $H(z)$ at low redshifts, affecting the distance to last scattering, i.e., $r_*$ in Eq.~\eqref{eq:ls-distance}.
Early-type solutions, on the other hand, change the sound horizon $r_s$ in Eq.~\eqref{eq:ls-soundhor}.
This can take the form of an increase in $H(z)$, for instance from early dark energy \citep{early-sol19ede,early-sol20edenu,early-sol19rock}.
In our case, however, it is through altering recombination.
Clumping changes the $z_*\left( \omega_r,\omega_b,\omega_m \right)$ function, which lowers $r_s$ (at fixed $\omega_i$); so to keep the same observed $\theta_s$, the comoving distance $r_*$ must be reduced, yielding higher $H_0$.

\subsection{Silk damping}

Another important phenomenon closely related to recombination physics is Silk diffusion damping \citep{silk-damping}. 
Photons perform a random walk with nonzero mean free path, which smooths their perturbations, making these decay as time passes.
A mode with wave number $k$ is suppressed by a factor $\mathcal{D}\left( k \right)$, which can be approximated as
\begin{equation}
\mathcal{D}\left( k \right) = \int_0^{\eta_0} d\eta \, g(\eta) \exp\left\{ -\left[ k/k_D \left( \eta \right) \right]^2 \right\}, \label{eq:damping-suppression} 
\end{equation}
where the effective diffusion scale is
\begin{equation}
k_D^{-2}\left( \eta \right) = \frac16 \int_0^{\eta} d\eta' \frac1{\dot\tau} \frac{R^2+16\left( 1+R \right)/15}{\left( 1+R \right)^2}, \label{eq:damping-scale}
\end{equation}
with $R=3\rho_b/4\rho_\gamma=\left(3\omega_b/4\omega_\gamma\right)\left(1+z\right)^{-1}$ \citep{cmb-damping-tail}, where $\rho_b$ and $\rho_\gamma$ are the physical energy densities of baryons and photons respectively.

As the visibility function is peaked around recombination ($\eta=\eta_*$) and normalized ($\int_0^{\eta_0}d\eta\, g\left(\eta\right)=1$), the damping factor can be approximated by taking out the exponential at peak out of the integral in Eq.~\eqref{eq:damping-suppression}, yielding $\mathcal{D}\left( k \right) \approx \exp\left\{ -\left[ k/k_D \left( \eta_* \right) \right]^2 \right\}$.
This introduces a new length scale into the problem: $r_D=2\pi/k_D$, which has different parameter dependence from $r_S$.

In the simplest case of a constant sound speed, $r_S$ scales as $c_S \eta_*= c_S \int_{z_*}^\infty dz/H(z)$, which is the distance a sound wave can travel up to recombination, 
whereas $r_D$ scales roughly as $\left[\int_0^{\eta_*} d\eta/(1+z)^2\right]^{1/2}=\left[\int_{z_*}^\infty dzH^{-1}(z)(1+z)^{-2}\right]^{1/2}$, as the comoving mean free path $\dot\tau^{-1}$ scales intrinsically as $(1+z)^{-2}$ [Eq.~\eqref{eq:inv-mfp-comov}]. 
As a consequence, the two scales will react differently to changes in the recombination history.
In particular, the damping scale receives a larger contribution from lower redshifts, near $z_*$, so it is more sensitive to the recombination profile.

Small-scale clumping shifts recombination to earlier times, and with it the peak of $g$,
as we showed in Fig.~\ref{fig:M3demo-rec} for our M3 model. 
This will alter the damping scale relative to the sound horizon.

\begin{figure}[ht!]
\includegraphics[width=\columnwidth]{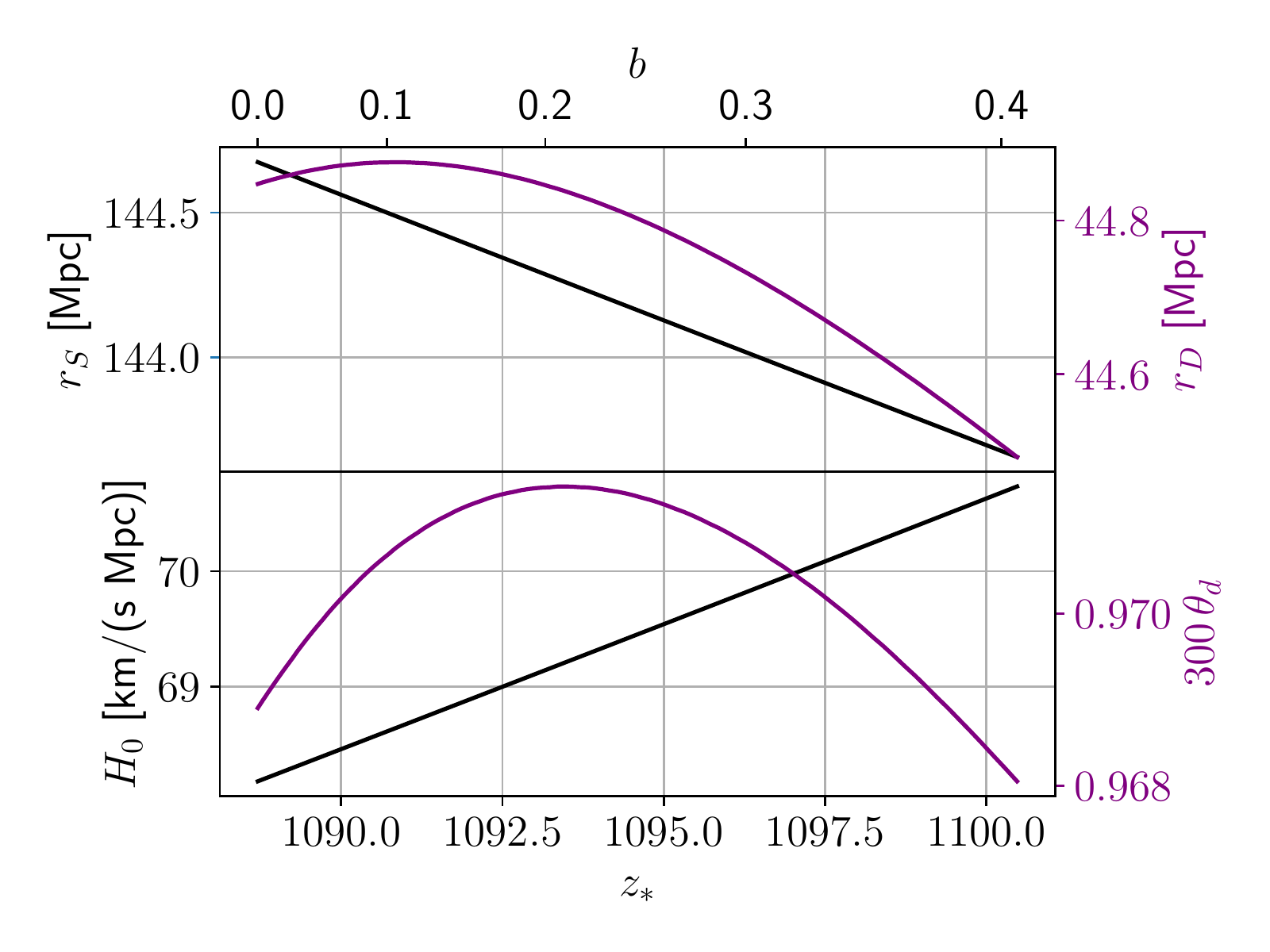}
\caption{Effects of clumping with a fixed cosmology ($\theta_s$, $\omega_b$, $\omega_\mathrm{cdm}$, $A_s$, $n_s$, $\tau_\mathrm{reio}$), in all cases with $f_0=2/3$ and $\left| \delta_+/\delta_- \right|=4/3$.
Top panel: the sound horizon $r_S$ (black) decreases with increasing clumping $b$ or recombination redshift $z_*$, whereas the damping scale $r_D$ (magenta) has a different and more complex behavior.
Bottom panel: $H_0$ increases (as distance to last scattering $r_*$ decreases proportionally to $r_s$, and other cosmological parameters are fixed); the angular damping scale $\theta_d=r_D/r_*$ first increases and then decreases.}
\label{fig:clumping-correlations}
\end{figure}

To build intuition, we have run a sequence of models with increasing clumping but fixed cosmology (in terms of $\theta_s$, which is exquisitely measured by the CMB), keeping $f_0=2/3$ and $\left| \delta_+/\delta_- \right|=4/3$ in our M3 model. 
The effects are shown in Fig.~\ref{fig:clumping-correlations}.
The comoving damping scale changes differently from the sound horizon, as we reasoned before.
Similarly to $\theta_s=r_S/r_*$, we convert it to an angular scale $\theta_d=r_D/r_*$, which first increases with clumping and then decreases.

CMB fluctuations with higher multipoles $\ell$ are further Silk suppressed, and thus provide a better measurement of the damping scale $k_D$. 
Therefore, good precision in the CMB damping tail provides a strong test of clumping.

\begin{figure}[ht!]
\includegraphics[width=\columnwidth]{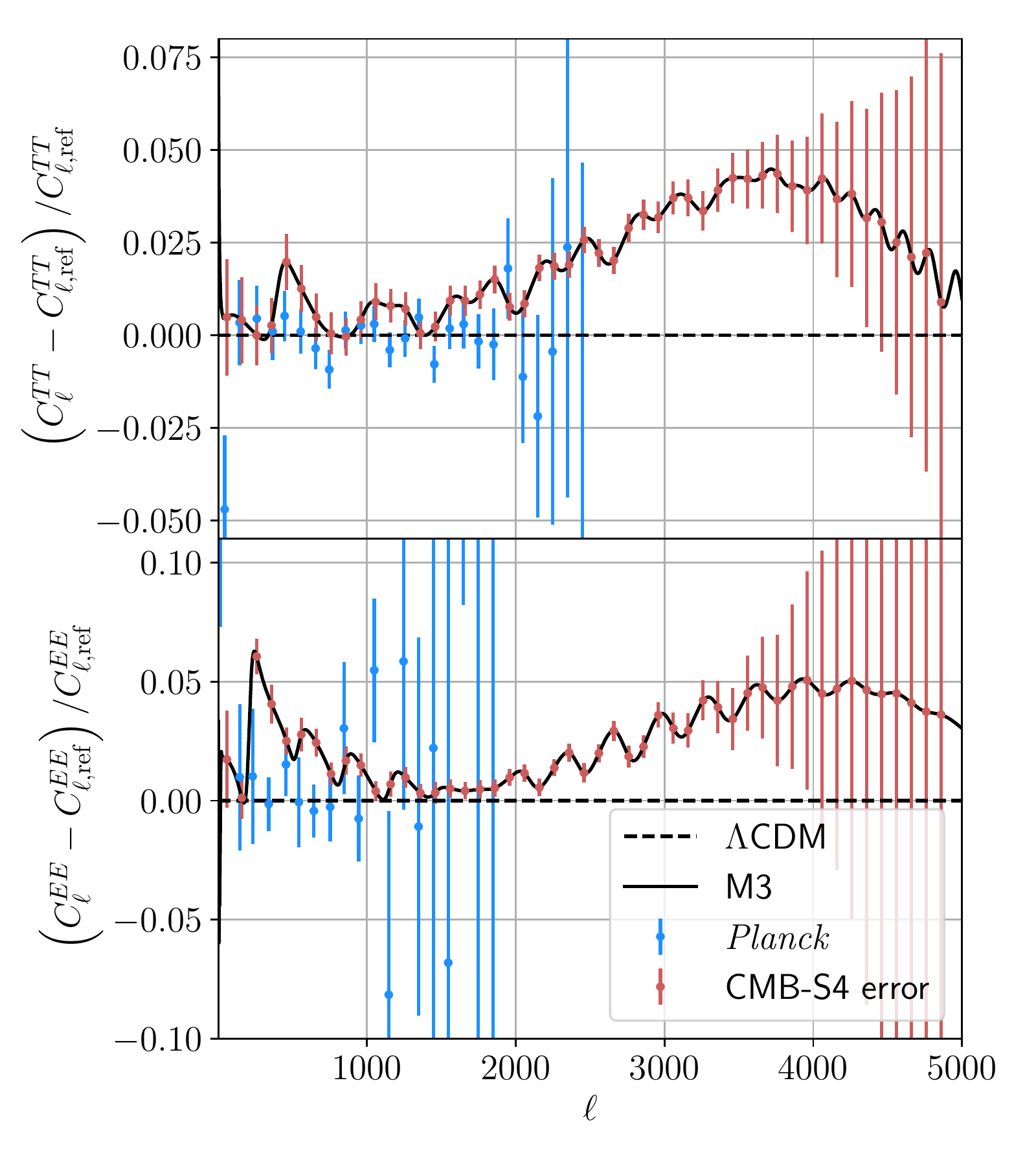}
\caption{
Demonstration of the effect of clumping on the CMB damping tail.
We define the $\Lambda$CDM prediction with standard recombination as the reference $C_{\ell,\rm ref}$, and compare our clumping model against it.
The black dashed line shows a clumping case, with M3 parameters $\delta_-=-0.9$, $\delta_+=5/3$, and $f_0=1/3$ (giving $b=1$). 
The cosmological parameters ($\theta_s$, $\omega_b$, $\omega_\mathrm{cdm}$, $A_s$, $n_s$, $\tau_\mathrm{reio}$) are fixed to the {\it Planck} best fit. 
Cyan points represent the binned {\it Planck} data, whereas red correspond to CMB-S4 forecasted error bars, both binned with $\Delta \ell=100$.}
\label{fig:M3demo-damping}
\end{figure}

We illustrate this in Fig.~\ref{fig:M3demo-damping}, where we show the relative difference between CMB power spectra, both temperature ($TT$) and polarization ($EE$) from M3 and from $\Lambda$CDM (with standard recombination), given the same cosmological parameters.
In particular, the same sound horizon angular scale $\theta_s$ ensures that the acoustic oscillations are in phase with one another, otherwise there would be a large oscillating difference between the power spectra.
The gradual deviation on smaller scales ($\ell\gtrsim 1500$ in temperature and $\ell\gtrsim 2000$ in polarization) is caused by the damping scale difference.
There are also smaller wiggles in the relative difference, which are caused by the change in duration of last scattering \citep{last-scattering-duration}.
However, they are less significant than the smoother trend.

We also show the {\it Planck} measurements and forecasted CMB-S4 binned errors (including both instrumental noise and cosmic variance).
Current measurements, from {\it Planck}, are scattered around zero, so that it is not easy to tell by eye how strongly this particular clumping configuration is disfavored. 
However, upon evaluating the likelihood we find $\Delta\chi^2_{Planck}=93$, most of it coming from high-$\ell$ ($\ell\ge 30$) $TT,TE,EE$ ($\Delta\chi^2_{{\rm high}\,\ell\,TTTEEE}=91$).
The difference is high because in this example we have fixed the rest of cosmological parameters.
As we will show later (Sec.~\ref{sec:planck-damping-scale}, Fig.~\ref{fig:bestfits-damping}), by shifting the cosmological parameters, M3 is able to fit the $\ell\lesssim 1000$ region very well, though at higher $\ell$ it diverges due to a difference in the damping scale.
With CMB-S4 errors, however, the difference induced by $b\sim 1$ clumping is clearly many sigmas in several dozens of bins both in temperature ($TT$) and polarization ($EE$), showing that more precise damping tail measurements will be able to distinguish the presence of significant clumping.
For this same example we find $\Delta\chi^2_{\rm CMB-S4}\approx 1350$, much higher than for {\it Planck}.

\section{Results with {\it Planck} 2018}
\label{sec:planck}

In this section we apply the M3 model to {\it Planck} 2018 data with the key goal of assessing how it alleviates the Hubble tension.
We also perform model comparison ($\Lambda$CDM with standard recombination versus M3) and discuss the compatibility with LSS measurements.

Our CMB datasets are low-$\ell$ $TT,EE$, binned nuisance-marginalized high-$\ell$ $TT,TE,EE$ \citep{planck_ttee} and lensing \citep{planck_lensing} power spectra.
We also consider the Hubble constant measurement from the SH0ES Collaboration: $H_0=\left( 73.2\pm 1.3 \right)$ km s$^{-1}$ Mpc$^{-1}$ \citep{SH0ES}.

We use the Cobaya framework \citep{cobaya} with the Polychord nested sampler \citep{polychord1,polychord2} for evidences (needed to compute the Bayes factors) and posteriors, and the Py-BOBYQA minimizer \citep{pybobyqa1,pybobyqa2,pybobyqa3} for best-fit determinations. Plots are made with anesthetic \citep{anesthetic} and GetDist \citep{getdist}.

\begin{table}[ht!]
\centering
\begin{tabular}{|c|cc|}
\hline
 & Prior & Range \\
\hline
$-\delta_-$ & Log-uniform & $[10^{-5},0.955]$ \\
$\left| \delta_+/\delta_- \right|$ & Log-uniform & $[0.1,10]$ \\
$f_0$ & Uniform & $[0,1]$ \\
\hline
$\ln(10^{10} A_s)$ & Uniform & $[2.55,3.55]$ \\
$n_s$ & Uniform & $[0.9,1.05]$ \\
$100\theta_s$ & Uniform & $[0.95,1.15]$ \\
$\Omega_b h^2$ & Uniform & $[0.02,0.025]$ \\
$\Omega_\mathrm{cdm} h^2$ & Uniform & $[0.1,0.15]$ \\
$\tau_\mathrm{reio}$ & Uniform & $[0.01,0.2]$ \\
\hline
$\Omega_K$ & Fixed & 0 \\
$m_\nu$ [eV] & Fixed & 0 \\
\hline
$A_{planck}$ ($y_{cal}$) & Normal & $1\pm 0.0025$ \\
\hline
\end{tabular}
\caption{Parameters used in this work and their priors.
$\delta_\pm$ and $f_0$ are only for M3 model, $A_{planck}$ ($y_{cal}$)---for {\it Planck} likelihoods.}
\label{tab:params}
\end{table}

Our parameters and priors are described in Table \ref{tab:params}.
It is important to state that we assumed massless neutrinos throughout the sampling to save computing time, as massive neutrinos slow the Boltzmann solver by a factor of $\sim 10$. 
This assumption shifts upwards the inferred $H_0$ values, but it does not affect the changes introduced by M3 (relative to $\Lambda$CDM with standard recombination) in a meaningful way, as we demonstrate in Appendix \ref{sec:justify-mnu}.

Full contour plots for the M3 clumping model are presented in Appendix \ref{sec:fullcontours}; here we will focus on particular important subspaces.

\subsection{\texorpdfstring{$H_0$}{H0} and model comparison}

\begin{figure}[ht!]
\includegraphics[width=\columnwidth]{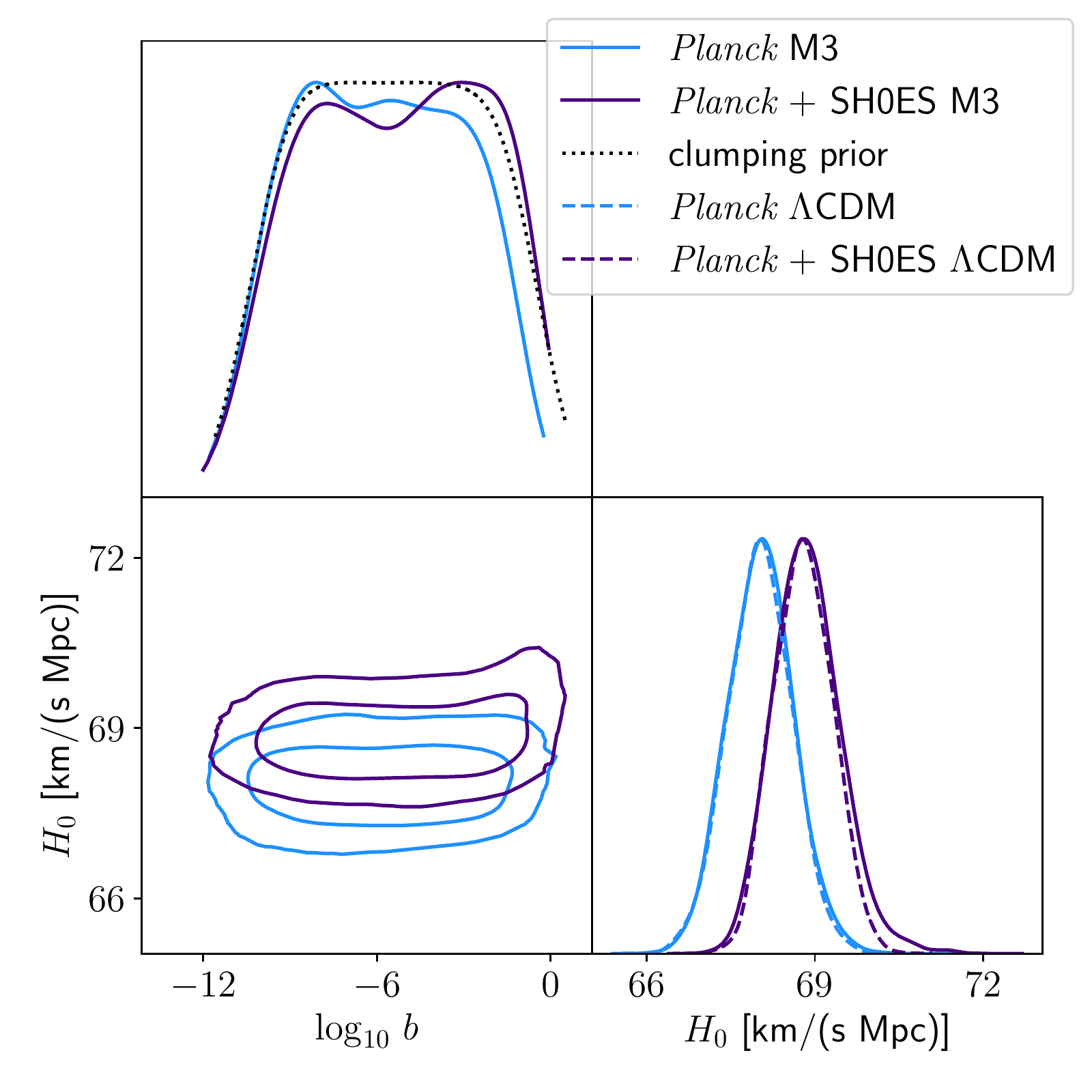}
\caption{Posteriors for $H_0$ and the clumping parameter $b$ of our M3 model [see Eq.~\eqref{eq:b-def} for its definition].
We show 68\% and 95\% CL ellipses using {\it Planck} 2018 data alone and in combination with SH0ES, which slightly increases $H_0$ though does not prefer clumping.
We also show the $H_0$ posteriors from $\Lambda$CDM runs with standard recombination, as well as the prior on $b$ for comparison.
The 68\% C.L. intervals on clumping within M3 are $\log_{10}\,b = -5.9\pm 2.7$ for {\it Planck}, $\log_{10}\,b = -5.3^{+4.2}_{-3.7}$ for {\it Planck}+SH0ES (and $\log_{10}\,b = -5.5\pm 3.0$ for prior).
Neither {\it Planck} only nor {\it Planck}+SH0ES data prefer large clumping ($b\sim 1$).}
\label{fig:planck-b-H0}
\end{figure}

We show the 2D posterior for the clumping parameter $b$ and $H_0$ in Fig.~\ref{fig:planck-b-H0}. {\it Planck}-only data show no noticeable change in $H_0$ compared to $\Lambda$CDM with standard recombination, 
and some preference against high clumping compared to the prior.
Adding a direct $H_0$ measurement, however, creates a weak preference for high clumping and high $H_0$.
Because most of the posterior weight for the clumping parameter $b$ is below unity, we observe almost no correlation between $b$ and $H_0$.

\begin{figure}[ht!]
\includegraphics[width=\columnwidth]{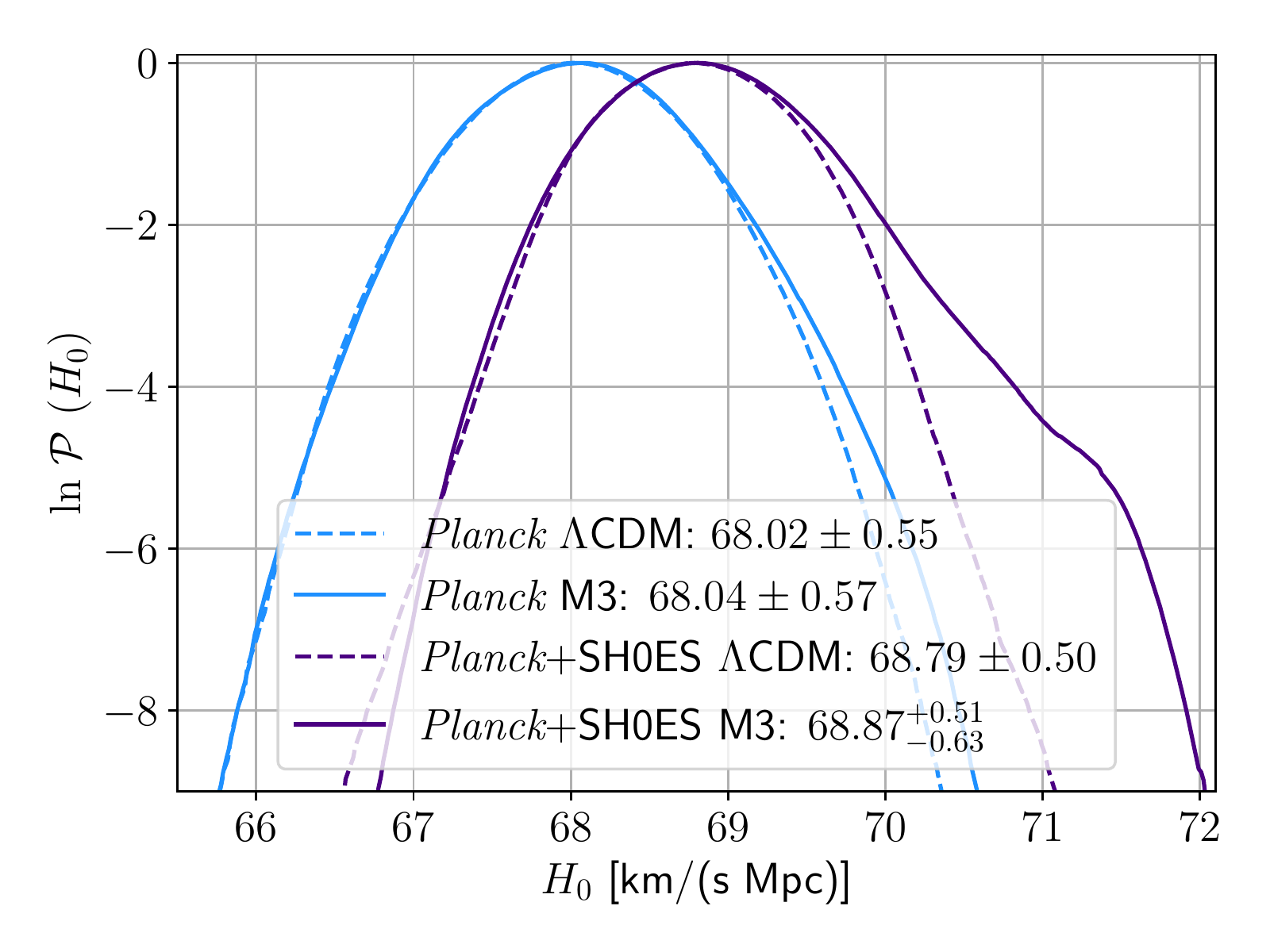}
\caption{Posterior of $H_0$ inferred from {\it Planck} data, without and with SH0ES, which shifts the $H_0$ values upward as expected.
The legend shows 68\% C.L. intervals.}
\label{fig:H0-planck}
\end{figure}

In order to explore the tail of the $H_0$ posterior distribution, we show it in Fig.~\ref{fig:H0-planck}.
With {\it Planck}-only data, M3 allows for a weak bump towards higher $H_0$ (compared to standard recombination), whereas the mean is not significantly shifted.
For {\it Planck} data combined with SH0ES, the bump at higher $H_0$ for M3 is stronger, given the additional pull from direct $H_0$ measurement, though the shift in the mean is still not significant [$\Delta H_0=0.08$ km/(s Mpc)].

\begin{table}[ht!]
\centering
\begin{tabular}{|c|c|c|}
\hline
 & $\Delta \chi^2_{best}$ & $\log_{10} K$ \\
\hline
{\it Planck} & 0 & $0.04\pm 0.15$ \\
\hline
{\it Planck}+SH0ES & 5 & $0.14\pm 0.19$ \\
\hline
\end{tabular}
\caption{Model comparison between M3 and $\Lambda$CDM by best-fit $\chi^2$ difference and Bayes factor $K$ between M3 and $\Lambda$CDM with {\it Planck} 2018 data.
$\chi^2$ differences rounded to integers because of uncertainty in the minimizer output.
M3 does not fit {\it Planck} data alone better than $\Lambda$CDM.
With SH0ES, M3 allows for slightly better agreement.
Bayes factors $-0.5\lesssim\log_{10}K\lesssim 0.5$ show no clear preference between models~\citep{bayes-factors}.}
\label{tab:model-comparison-current}
\end{table}

We show the best-fit $\chi^2$ differences and Bayes factors $K$ between M3 and $\Lambda$CDM models in Table \ref{tab:model-comparison-current}.
We have not found a better M3 fit to {\it Planck} 2018 data alone, compared to $\Lambda$CDM with standard recombination.
M3 is more successful than $\Lambda$CDM when considering {\it Planck}+SH0ES, but $\Delta\chi^2\approx 5$ can not justify three extra parameters.
The Bayes factor $K$ in both cases is consistent with 1 (within $<1\sigma$), meaning no preference to either model.
The Bayes factor is equal to the ratio of marginalized posterior probabilities of the models if they are assumed equally probable {\it a priori}; more generally, posterior probabilities ratio is given by the Bayes factor times prior probabilities ratio \citep{bayes-factors}.
The best-fit parameters are presented in Appendix \ref{sec:bestfits}, Table \ref{tab:bestfits-planck}.

We conclude that M3 is neither supported by the data nor rejected.
This is probably not surprising, since {\it Planck} data are fit well by $\Lambda$CDM with standard recombination and with an $\approx 4\sigma$ tension between {\it Planck} and SH0ES it is challenging to get a detection in a three-parameter model.
One could get more support for M3 by considering additional $H_0$ data.
Here, however, we will limit ourselves to the SH0ES measurement.

\subsection{Low-\texorpdfstring{$\ell$}{l} {\it Planck} data analysis}

In order to build intuition, we now check whether it is the damping tail that prevents {\it Planck} 2018 data from preferring M3.
For that, we perform an analysis with only the $\ell<1000$ multipoles. 
This range of scales is chosen to determine the sound horizon angular scale, though not the damping tail.

\begin{figure}[ht!]
\includegraphics[width=\columnwidth]{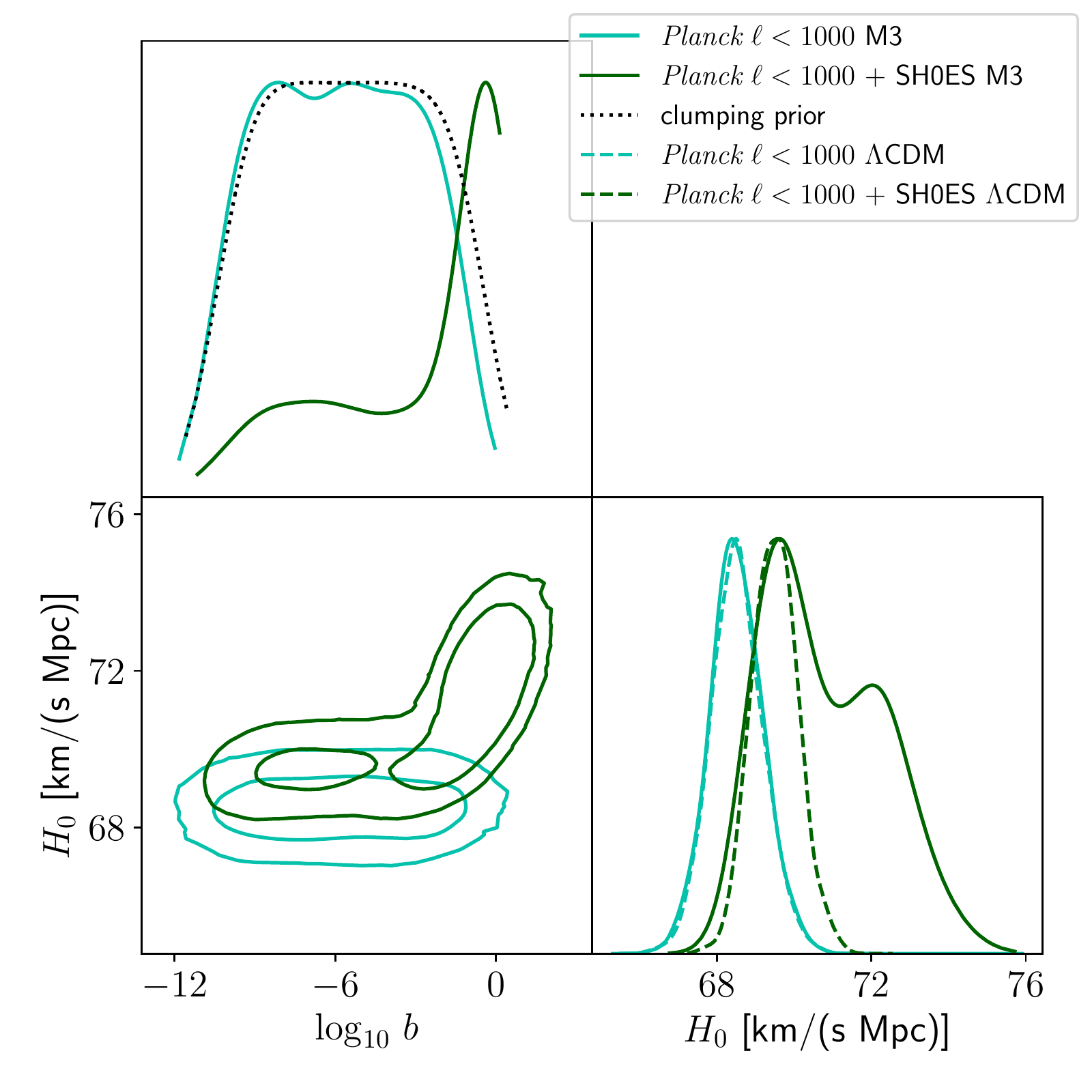}
\caption{Confidence regions for clumping $b$ versus $H_0$ from {\it Planck} 2018 $\ell<1000$ data.
The absence of damping tail information from the high-$\ell$ data allows significant clumping, and thus larger $H_0$.
The 68\% C.L. intervals on clumping within M3 are $\log_{10}\,b = -5.9\pm 2.8$ for {\it Planck} $\ell<1000$, $\log_{10}\,b = -2.8^{+2.9}_{-6.3}$ for {\it Planck} $\ell<1000$ + SH0ES (and $\log_{10}\,b = -5.5\pm 3.0$ for prior).
Within M3, $H_0 = (68.51\pm 0.68)$ km/(s Mpc) for {\it Planck} $\ell<1000$, $H_0 = (70.8\pm 1.5)$ km/(s Mpc) for {\it Planck} $\ell<1000$ + SH0ES.
The $H_0$ posteriors from $\Lambda$CDM runs are also shown, which do not reach the high $H_0$ values available to M3.}
\label{fig:planck-lowl-b-H0}
\end{figure}

The corresponding 2D posteriors for $b$ and $H_0$ are plotted in Fig.~\ref{fig:planck-lowl-b-H0}.
Using {\it Planck} $\ell<1000$, we find no significant deviation from standard recombination, as the clumping posterior is only slightly shifted from the prior (towards lower values).
However, with the additional pull from SH0ES, strong clumping $b\sim 1$ is preferred, and we see a significant bump towards higher $H_0$.
The best-fit parameters are presented in Appendix \ref{sec:bestfits}, Table \ref{tab:bestfits-planck-lowl}.

\begin{table}[ht!]
\centering
\begin{tabular}{|c|c|c|}
\hline
 & $\Delta \chi^2_{best}$ & $\log_{10} K$ \\
\hline
{\it Planck} $\ell<1000$ & 0 & $-0.09\pm 0.18$ \\
\hline
{\it Planck} $\ell<1000$+SH0ES & 11 & $0.21\pm 0.20$ \\
\hline
\end{tabular}
\caption{Model comparison with {\it Planck} $\ell<1000$ data.
$\chi^2$ differences are rounded to integers because of uncertainty in the minimizer output.
M3 still does not fit the CMB alone better than $\Lambda$CDM, though the joint fit to CMB and SH0ES is improved further than with full {\it Planck}.
The Bayes factors $K$ show no preference between models, as for full Planck in Table \ref{tab:model-comparison-current}.}
\label{tab:model-comparison-lowl}
\end{table}

In order to determine whether clumping provides a better fit in this case, we perform model comparison by two methods: best-fit $\chi^2$ and Bayes factor $K$, and show the results in Table \ref{tab:model-comparison-lowl}. 
We still do not find M3 to fit CMB data notably better than the standard model, with the addition of SH0ES improvement is more significant ($\Delta\chi^2_{best}=11$) than in the case of full {\it Planck} and SH0ES ($\Delta\chi^2_{best}=5$).
Bayes factors $K$ are still consistent with one, telling no clear preference between the models.
This shows that current {\it Planck} data do not prefer clumping as the solution to the $H_0$ tension, even without its damping tail.

\subsection{Damping scale}
\label{sec:planck-damping-scale}

As we have shown in the previous subsection, considering only $\ell<1000$ multipoles from the {\it Planck} data opens more room for clumping than using the full data.
We posit that the key source of constraints on clumping is the damping-tail information contained in higher-$\ell$ multipoles.
So now we consider how the damping tail varies within our M3 model.

\begin{figure}[ht!]
\includegraphics[width=\columnwidth]{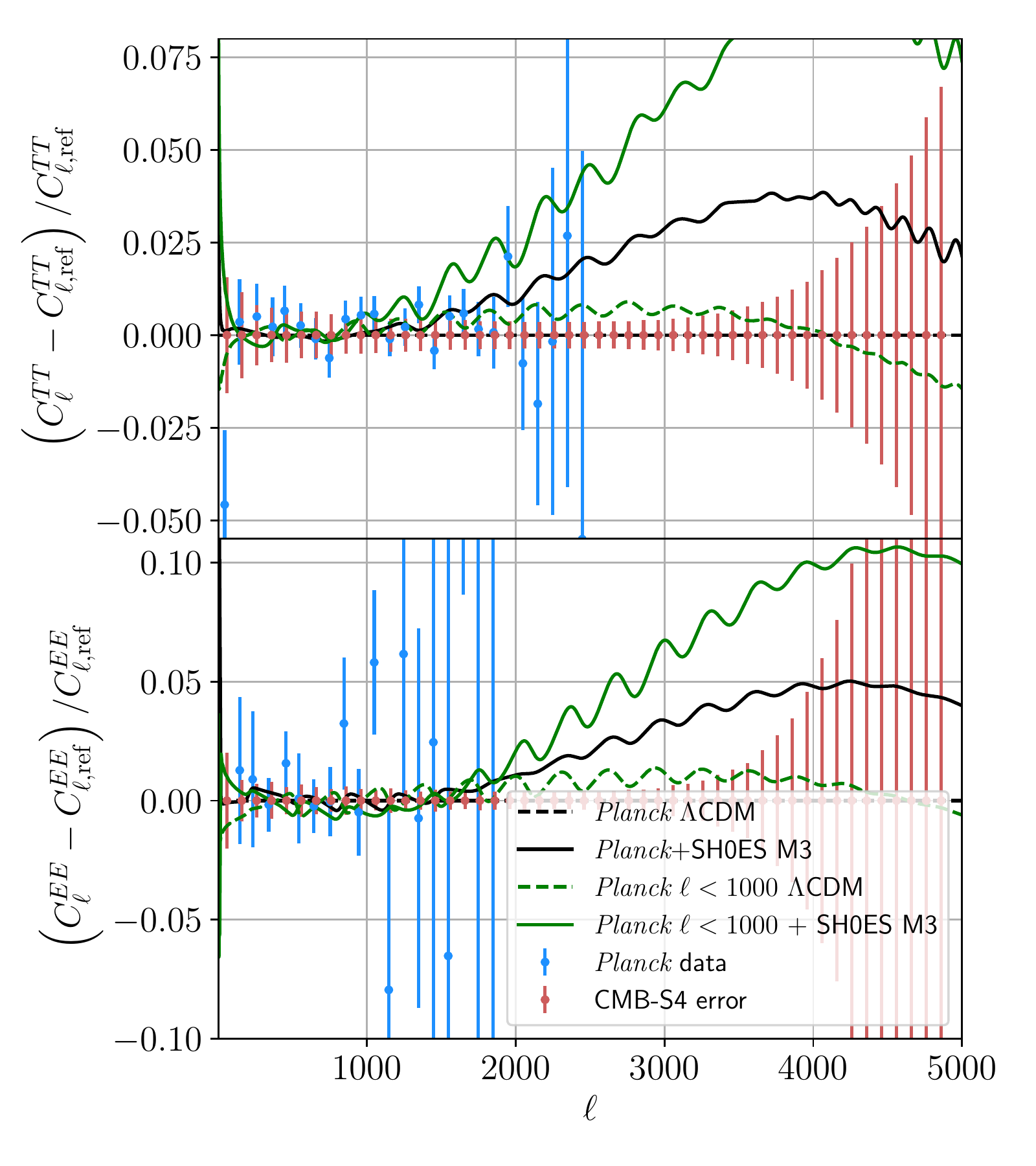}
\caption{Relative difference in CMB temperature $TT$ and polarization $EE$ power spectra between our best fits (Appendix \ref{sec:bestfits}, Tables \ref{tab:bestfits-planck}, \ref{tab:bestfits-planck-lowl}), with the $\Lambda$CDM best fit to {\it Planck} taken as reference.
Differences are small in the low-$\ell$ regime, but grow at smaller scales.
{\it Planck} data can barely distinguish them, but CMB-S4 will do so very clearly.}
\label{fig:bestfits-damping}
\end{figure}

First, we plot relative difference in $TT,EE$ power spectra between the best fits in Fig.~\ref{fig:bestfits-damping}.
Unlike in Fig.~\ref{fig:M3demo-damping}, where the cosmology was kept fixed, producing significant differences, here all the models manage to shift the parameters to fit the low-$\ell$ data.
However, they diverge significantly in the damping tail ($\ell\gtrsim 1500$), and CMB-S4 will be able to measure such deviations.
More precisely, the $\chi^2$ difference between {\it Planck} $\Lambda$CDM and {\it Planck}+SH0ES M3 best fits from {\it Planck} temperature, polarization and lensing is only $\Delta\chi^2_{Planck}\approx 4$, while for CMB-S4 precision in temperature and polarization it can be as high as $\Delta\chi^2_{\rm CMB-S4}\approx 1030$ (if one assumes the {\it Planck} $\Lambda$CDM best fit as fiducial).

In order to build intuition, we formulate the damping-tail constraints in terms of the comoving damping scale $r_D$ and its angular analog $\theta_d=r_D/r_*$.
We postprocessed the {\it Planck} runs to get this information from CLASS.

\begin{figure}[ht!]
\includegraphics[width=\columnwidth]{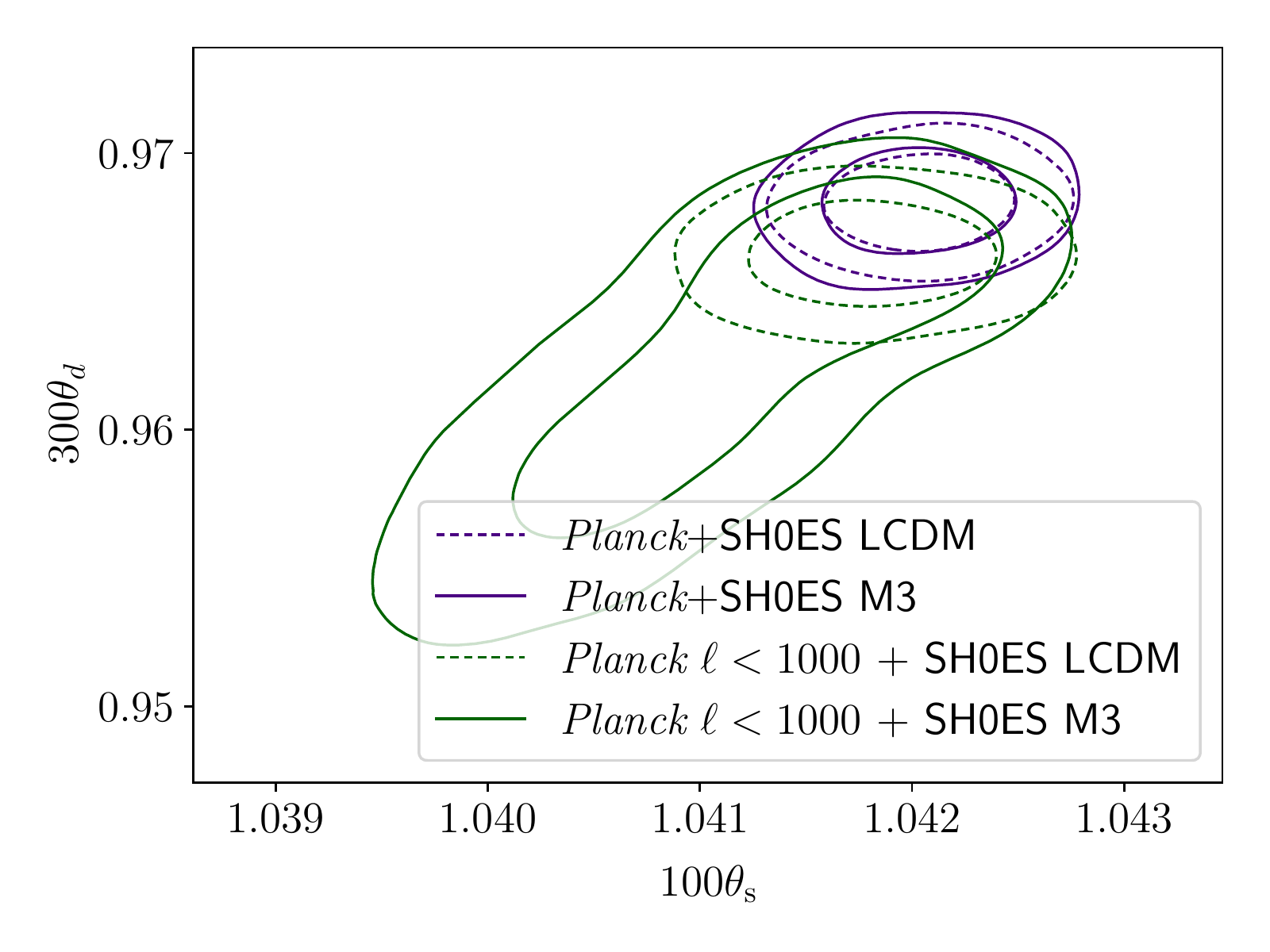}
\caption{Sound-horizon vs. damping angular scales [both multiplied by a prefactor to make them $\mathcal O(1)$] for {\it Planck} 2018 + SH0ES runs.
The M3 and $\Lambda$CDM contours are quite similar for full {\it Planck}+SH0ES (purple).
On the contrary, for {\it Planck} $\ell<1000$+SH0ES (green), the M3 posterior is significantly extended to lower values of both $\theta_s$ and $\theta_d$, compared to $\Lambda$CDM.}
\label{fig:planck-theta_s-d}
\end{figure}

Figure \ref{fig:planck-theta_s-d} presents the sound-horizon vs. damping angular scales for our {\it Planck}+SH0ES runs. 
Within $\Lambda$CDM both angular scales are well measured, both for the full {\it Planck} data as well as with the $\ell<1000$ modes only.
The M3 contours, however, extend to lower values of both $\theta_d$ and $\theta_s$ compared to $\Lambda$CDM. 
With full {\it Planck} the shift is not significant, while using only $\ell<1000$ allows larger deviations to that region. 
With {\it Planck} $\ell<1000$ + SH0ES, the error bar of $\theta_d$ within M3 is by a factor of 3 wider than within $\Lambda$CDM.

\begin{figure}[ht!]
\includegraphics[width=\columnwidth]{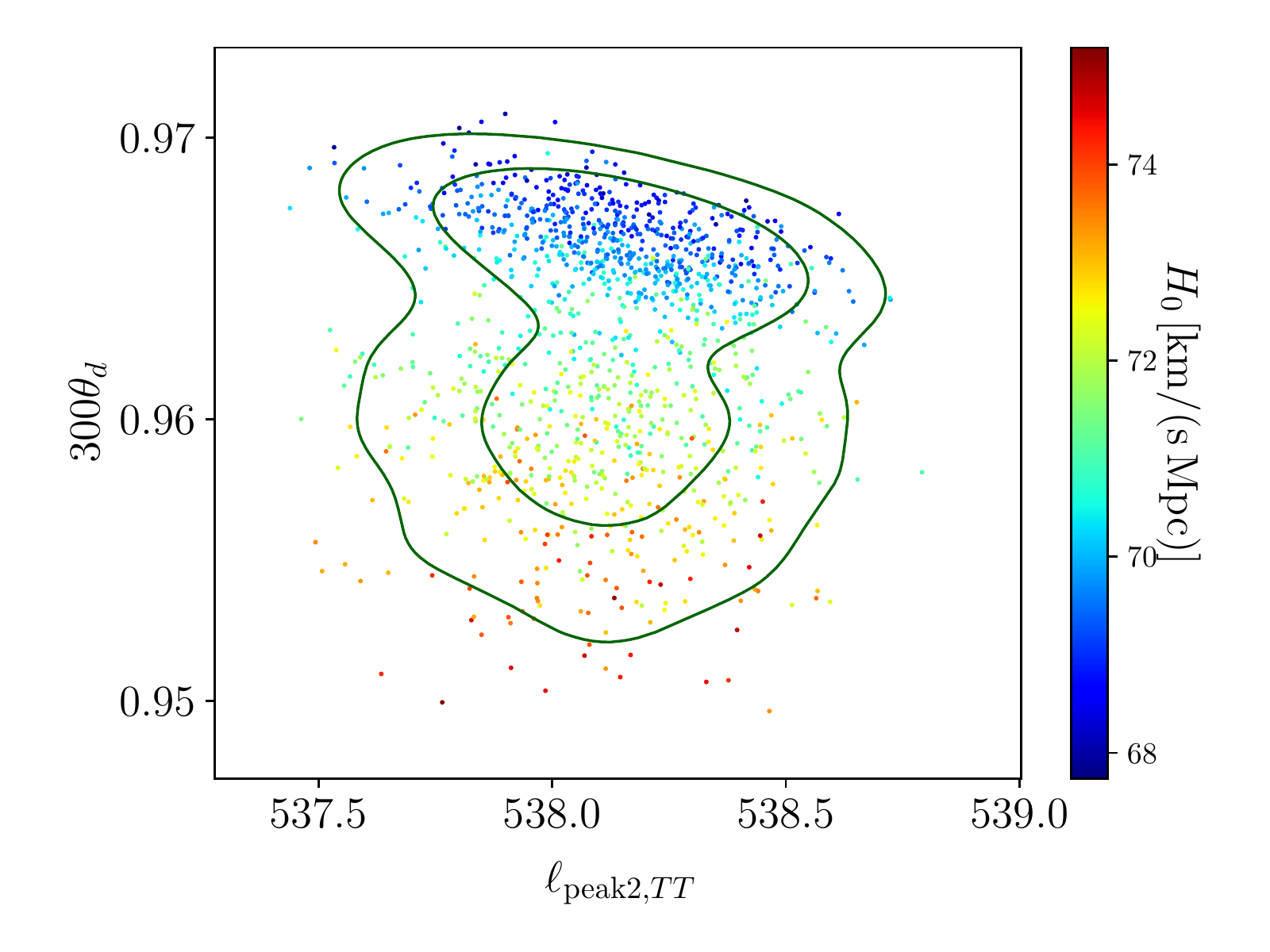}
\caption{Similar to Fig.~\ref{fig:planck-theta_s-d}, but we replace $\theta_s$ by position of the second peak in $TT$ power spectrum $\ell_{{\rm peak2},TT}$ and color different points by their value of $H_0$, for our {\it Planck} $\ell<1000$ + SH0ES M3 run.
Within the M3 peak positions stay almost same regardless of $\theta_d$.
This proves that $\ell<1000$ multipoles from {\it Planck} give strong enough constraints on acoustic peaks.
In this plane, $H_0$ increases for lower values of both $\theta_d$ and $\ell_{\rm peak}$.
However, the peak positions are measured with $\sim 0.1\%$ precision and do not allow to vary $H_0$ significantly with other parameters fixed.
This implies that a change in the angular damping scale is necessary to infer higher $H_0$ from the CMB.}
\label{fig:planck-SH0ES-lowl-ell_peak2-theta_d-H0}
\end{figure}

We note that the error bar of $\theta_s$ is widened similarly.
However, the positions of acoustic peaks are not determined exactly by sound horizon angular scale alone.
For example, stronger damping would shift the power spectrum maxima to slightly lower $\ell$.
Acoustic peak positions may also be affected by fine changes in visibility function shape introduced by clumping.
Such small changes might be important, since the characteristic differences in sound horizon scales in our runs are only $\sim 0.1\%$.
To assess this, we calculate the positions of first three peaks in $TT$ power spectrum given by CLASS [by fitting Gaussians to $\mathcal{D}_\ell=\ell(\ell+1)C_\ell/2\pi$, as {\it Planck} Collaboration \citep{planck_overview}].
We find that M3 keeps peaks at the same positions as $\Lambda$CDM, especially the second one.
Therefore we plot second $TT$ peak position instead of $\theta_s$ in Fig.~\ref{fig:planck-SH0ES-lowl-ell_peak2-theta_d-H0}.

Figure \ref{fig:planck-SH0ES-lowl-ell_peak2-theta_d-H0} also shows how $H_0$ changes in the $\ell_{{\rm peak2},TT}$-$\theta_d$ plane, using {\it Planck} 2018 $\ell<1000$ and SH0ES data. 
The main trend is that $H_0$ increases for smaller values of $\theta_d$.
Therefore, lowering the angular damping scale $\theta_d$ is necessary to infer higher $H_0$ from CMB.
But such change is disfavored by {\it Planck} damping tail data.
This agrees with our previous subsections, where full {\it Planck} + SH0ES did not show a preference for high clumping, and consequently did not exhibit significant change in $H_0$, while without $\ell\ge 1000$ multipoles the data allowed for both.

\subsection{\texorpdfstring{$S_8$}{S8} tension}

We now move to discuss whether clumping is compatible with large-scale structure data, which have not been considered in previous subsections.

A potentially interesting discrepancy between the CMB and LSS is the $S_8$ tension in the amplitude of matter fluctuations measured from the CMB and the LSS.
{\it Planck} 2018 reported $\Omega_m=0.315\pm 0.007$ and $S_8=0.831\pm 0.017$ \citep{planck_cosmo}, both of which are higher than found in DES-Y1: $\Omega_m=0.264^{+0.032}_{-0.019}$, $S_8=0.783^{+0.021}_{-0.025}$ \citep{DES-Y1}.
Earlier recombination (for instance due to small-scale baryon clumping) decreases both the $\Omega_m$ and $S_8$ values inferred from the CMB, and can therefore help to relieve the $S_8$ tension \citep{JP20}.

However, the new DES-Y3 results ($\Omega_m=0.339^{+0.032}_{-0.031}$, $S_8=0.776\pm0.017$ \citep{DES-Y3}) do not show a preference for lower values of $\Omega_m$, which makes clumping less favorable resolution.
A reanalysis of DES-Y1 data according to the DES-Y3 pipeline shifted the parameter estimates to $\Omega_m=0.303^{+0.034}_{-0.041}$, $S_8=0.747^{+0.027}_{-0.025}$ \citep{DES-Y3}, in better agreement with {\it Planck} on $\Omega_m$ but worse on $S_8$.
Other experiments also find a lower value of $S_8$ than {\it Planck} and only a small difference in $\Omega_m$ ($<1\sigma$),
including the
Kilo-Degree Survey (KiDS-1000, which reported $\Omega_m=0.305^{+0.010}_{-0.015}$, $S_8=0.766^{+0.020}_{-0.014}$ \citep{KiDS-1000}), 
unWISE galaxies (with Planck CMB lensing added, which obtained $\Omega_m=0.295\pm 0.017$, $S_8=0.776\pm 0.017$ \citep{unWISE-plancklensing}),
as well as an analysis of the growth of density perturbations from large-scale structure data (which yielded $\Omega_m=0.311^{+0.021}_{-0.028}$, $S_8=0.7769\pm 0.0095$ \citep{dens-perturb-growth}, see also \citep{sdss-boss-eft1,sdss-boss-eft2}).

\begin{figure*}[p!]
\includegraphics[width=\textwidth]{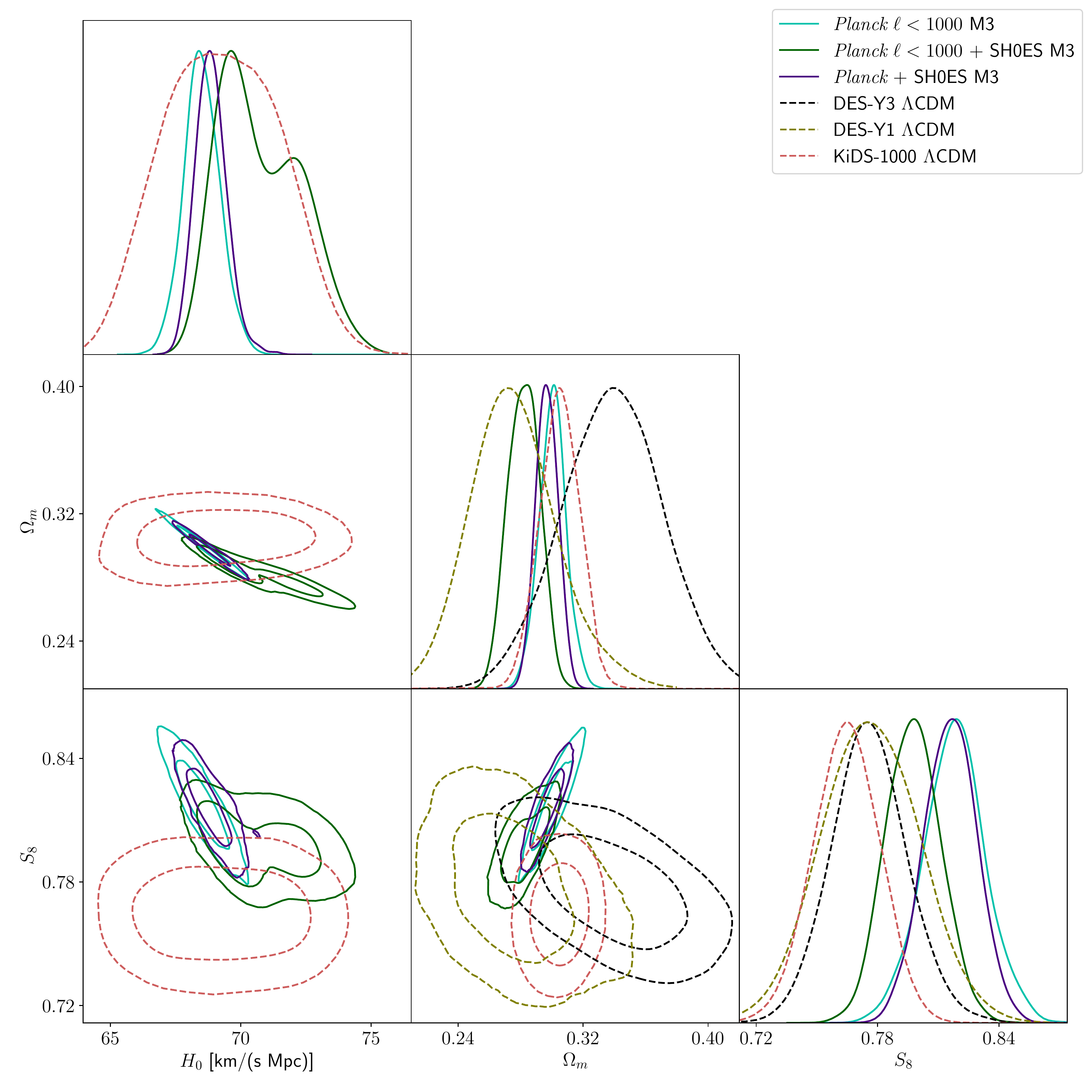}
\caption{Confidence regions and one-dimensional posteriors for $H_0$, matter fraction $\Omega_m$ and rescaled clustering amplitude $S_8$.
{\it Planck} $\ell<1000$ M3 has low clumping, therefore is close to $\Lambda$CDM and exhibits similar negative $H_0$--$\Omega_m$ and $H_0$--$S_8$ correlations, so increasing $H_0$ decreases both $\Omega_m$ and $S_8$.
{\it Planck} $\ell<1000$ + SH0ES M3 can reach high clumping (see Fig.~\ref{fig:planck-lowl-b-H0}), giving rise to even lower $\Omega_m$ and $S_8$.
High clumping is disfavored by the damping-tail data of full {\it Planck}+SH0ES, so the corresponding contours are closer to $\Lambda$CDM.
We show the confidence regions and posteriors from DES-Y3 \citep{DES-Y3}, DES-Y1 \citep{DES-Y1} and KiDS-1000 \citep{KiDS-1000} for reference.
DES data do not constrain $H_0$ significantly compared to their prior, which is flat in the range 55 to 91 km/(s Mpc).}
\label{fig:planck-combined-matter-clustering}
\end{figure*}

To study in detail how clumping interfaces with the $S_8$ tension, we show the one-dimensional posteriors and two-dimensional confidence ellipses for $H_0$, $\Omega_m$ and $S_8$ for a few selected runs in Fig.~\ref{fig:planck-combined-matter-clustering}.
We also show analogous posteriors and contours on $\Omega_m$ and $S_8$ from DES-Y3 \citep{DES-Y3}, DES-Y1 \citep{DES-Y1} and KiDS-1000 \citep{KiDS-1000} for comparison.
This figure shows that {\it Planck} $\ell<1000$ M3 prefers low clumping and therefore is closer to $\Lambda$CDM with standard recombination, exhibiting similar correlations between these three parameters.
Note that within $\Lambda$CDM there is a significant negative correlation between $H_0$ and $\Omega_m$ and a weaker negative correlation between $H_0$ and $S_8$.
Therefore increasing $H_0$ alone (for instance by coadding direct $H_0$ measurements) decreases both $\Omega_m$ and $S_8$.

The {\it Planck} $\ell<1000$ + SH0ES M3 confidence region exhibits high clumping and explores a different direction, to even lower $\Omega_m$ and $S_8$.
Finally, in full {\it Planck} with SH0ES the damping-tail data disfavor high clumping, making the contour close to the standard $\Lambda$CDM.

We note that a rigorous study of the $S_8$ tension within M3 would require a reanalysis of the LSS data, as clumping can introduce biases with respect to $\Lambda$CDM with standard recombination.
Such an analysis is beyond the scope of this work, and given that this tension is weaker than the $H_0$ one, we tentatively conclude that adding LSS data to our runs would not significantly change whether M3 is preferred.

\subsection{Baryon drag scale}

As advanced above, the addition of clumping changes the length of the sound horizon due to the nonstandard recombination.
This is important for the interpretation of baryon acoustic oscillations (BAO) in galaxy surveys at low $z$, so we now consider how the BAO standard ruler is affected by clumping.
The relevant distance is the drag scale $r_{\rm drag}$---the sound horizon at the drag epoch (when the baryon optical depth is 1).
The drag epoch occurs slightly later than last scattering, which makes the drag scale larger than the sound horizon $r_S$ at last scattering, albeit only marginally.

Standard-ruler BAO measurements constrain the combinations $d_M\left(z\right)/r_{\rm drag}$ and $H\left(z\right)r_{\rm drag}$ \citep{BAO-constraints}, where $d_M$ is the comoving angular-diameter distance.
In a flat $\Lambda$CDM cosmology,
\begin{equation}
    d_M = \int_0^z \frac{cdz}{H\left(z\right)},
\end{equation}
and the dominant contributors are the cosmological constant and the nonrelativistic matter, since for low $z$
$$ H\left(z\right)\approx H_0\left(1+\Omega_m\left[\left(1+z\right)^3-1\right]\right)^{1/2}. $$
For a constant $\Omega_m$, both $d_M\left(z\right)/r_{\rm drag}$ and $H\left(z\right)r_{\rm drag}$ depend only on $H_0r_{\rm drag}$.
On the CMB side, the sound-horizon angular scale $\theta_s$ is also proportional to $H_0r_{\rm drag}$ for fixed $\Omega_m$, since $r_{\rm drag}$ is very close to $r_S$ [see Eq.~\eqref{eq:ls-soundhor}].
The early integrated Sachs-Wolfe effect in the CMB determines the physical matter density $\omega_m=\Omega_mh^2$, which indeed stays roughly constant in our sampling.
Therefore, as $H_0$ increases, $\Omega_m$ decreases, which causes changes in $d_M\left(z\right)/r_{\rm drag}$ and $H\left(z\right)r_{\rm drag}$ at low redshift, compared to their high-redshift analog $\theta_s$ (effectively fixed by the CMB).
As a consequence, clumping models that fit the CMB develop a tension with BAO data (see \citet{sound-horizon-not-enough} for a broader discussion).

\begin{figure}[htp!]
\includegraphics[width=\columnwidth]{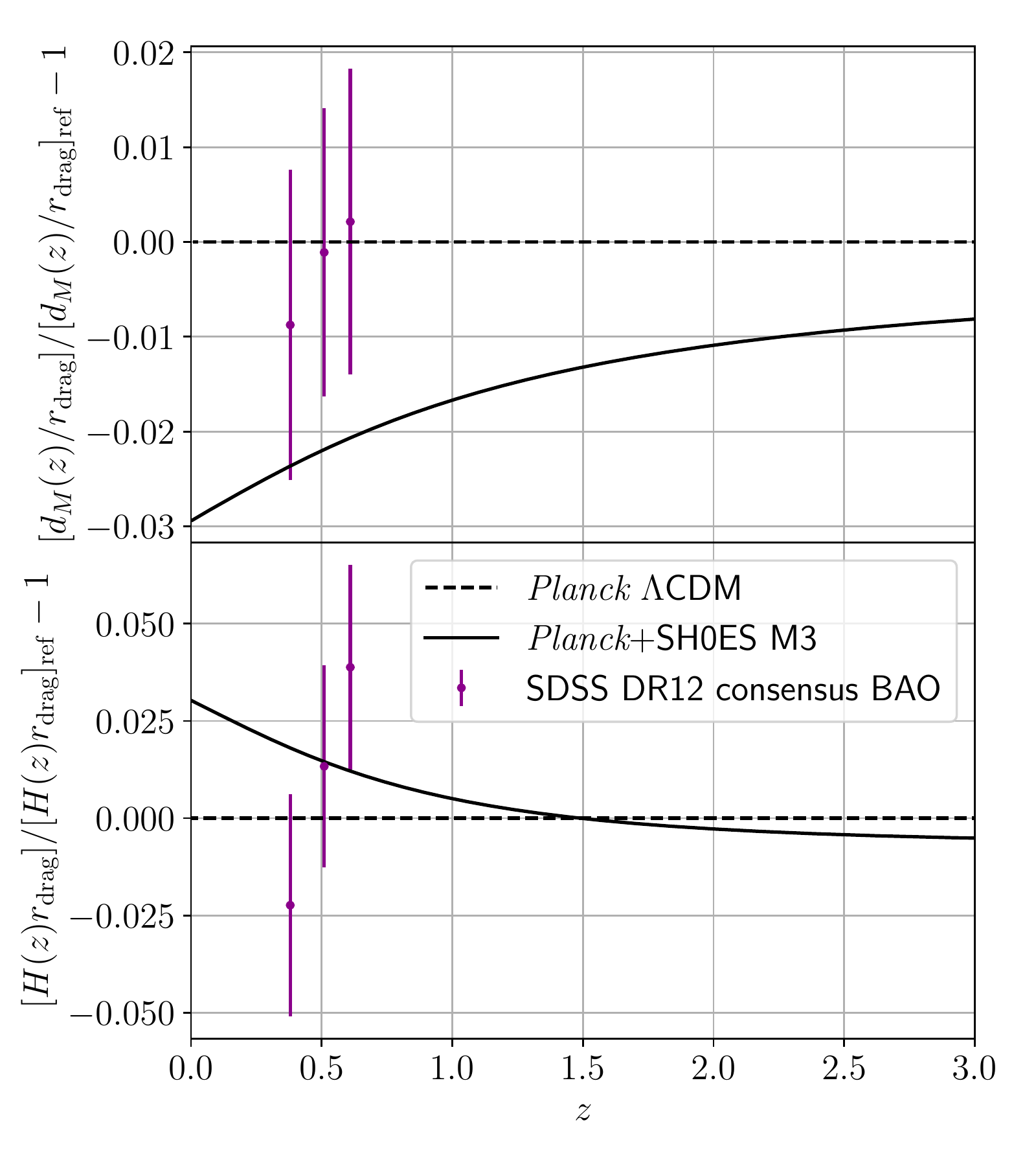}
\caption{Relative difference in the BAO distances---$d_M\left(z\right)/r_{\rm drag}$ and $H\left(z\right)r_{\rm drag}$,
for M3 best fit to {\it Planck}+SH0ES, compared to the $\Lambda$CDM best fit to {\it Planck} (see Appendix \ref{sec:bestfits}, Table \ref{tab:bestfits-planck} for exact parameters)
SDSS DR12 measurements \citep{BAO-constraints} are overlaid.
At higher redshifts, the relative difference in $d_M/r_{\rm drag}$ tends to 0 and in $Hr_{\rm drag}$---to a negative constant $\approx -0.01$.}
\label{fig:bestfits-BAO}
\end{figure}

As an example, in Fig.~\ref{fig:bestfits-BAO} we show the relative difference between these quantities for $\Lambda$CDM (best fit to {\it Planck}) and for M3 (best fit to {\it Planck}+SH0ES), and overlay the SDSS DR12 measurements \citep{BAO-constraints}.
By eye, we can tell that the {\it Planck}+SH0ES M3 best fit is mildly disfavored by the transversal [$d_M\left(z\right)/r_{\rm drag}$] BAO data, while for the radial [$H\left(z\right)r_{\rm drag}$] BAO the data scatter is too large to tell.
Using the full covariance matrix, we find $\Delta\chi^2_{\rm BAO}\approx-3.6$, indeed mildly disfavoring M3.

At higher redshifts, the relative difference in $d_M\left(z\right)/r_{\rm drag}$  tends to 0---so as to match $\approx 1/\theta_s$ to the CMB at recombination.
The $H\left(z\right)r_{\rm drag}$ relative difference, on the other hand, tends to a negative constant---since $\omega_m$ changes weakly, expansion rate at high redshifts is almost the same, so the difference is driven by the change in sound horizon (and thus drag scale), which is $\approx 1\%$ for the {\it Planck}+SH0ES M3 best fit compared to {\it Planck} $\Lambda$CDM best fit.

\begin{figure}[htp!]
\includegraphics[width=\columnwidth]{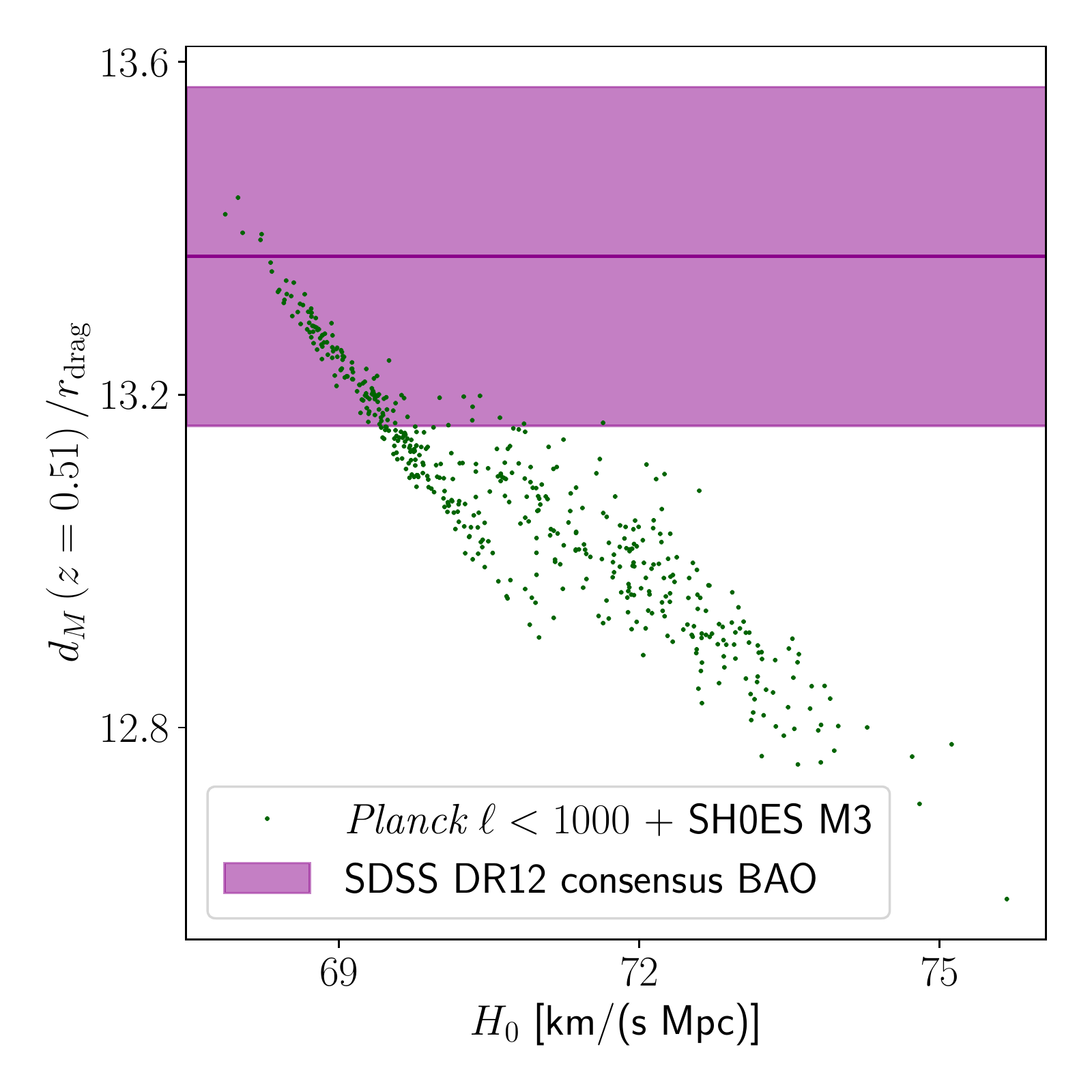}
\caption{Scatter plot of $d_M(z=0.51)/r_s$ versus $H_0$ for our {\it Planck} $\ell<1000$ + SH0ES M3 run, which exhibits the highest clumping.
We overplot the SDSS DR12 measurements at this redshift in magenta, where the semitransparent band shows its $1\sigma$ error bar.
Even though high clumping (in the right) produces higher $H_0$ than inferred from the $\Lambda$CDM degeneracy line (in the upper left), larger values of $H_0$ still cause tension between CMB and the BAO.}
\label{fig:BAO-scatter}
\end{figure}

While the M3 best fit to {\it Planck}+SH0ES is in mild tension with the current BAO measurements, that
does not necessarily prove that increasing $H_0$ in M3 is always disfavored by BAO data.
There remains a possibility that model parameters can be adjusted to accommodate the datasets and provide a better joint fit.
To assess this, we plot $d_M(z=0.51)/r_s$ versus $H_0$ for our {\it Planck} $\ell<1000$ + SH0ES M3 run in Fig.~\ref{fig:BAO-scatter}, overlaying the SDSS DR12 measurement.
The upper left dots have low clumping, so they follow the standard $\Lambda$CDM degeneracy direction.
The right dots, with high clumping, follow a different trend, but still develop more and more tension with SDSS for increasing $H_0$, as explained in \citet{sound-horizon-not-enough}.
We remind the reader that this figure shows only one of three SDSS D12 measurements, and the trend is similar in all, which makes the tension stronger.

Standard-ruler BAO data are being improved, and in the future it will become more decisive for or against the clumping model.
A notable example is the {\it Dark Energy Spectroscopic Instrument} (DESI), which is already operational.
DESI is expected to provide subpercent precision measurements of $d_M/r_{\rm drag}$ in seven bins for $0.65\le z\le 1.25$ \citep{DESI1}, which alone can give $\Delta\chi^2\approx 20$ between our best-fit models.
At higher redshifts, 21-cm data will provide a measurement of $H\, r_{\rm drag}$ to percent-level precision using on velocity-induced acoustic oscillations (VAOs)~\citep{HERA-VAO}.
Our clumping model predicts only a modest deviation of the radial $H\, r_{\rm drag}$, so transverse BAO measurements have more constraining power.

\begin{figure}[htp!]
\includegraphics[width=\columnwidth]{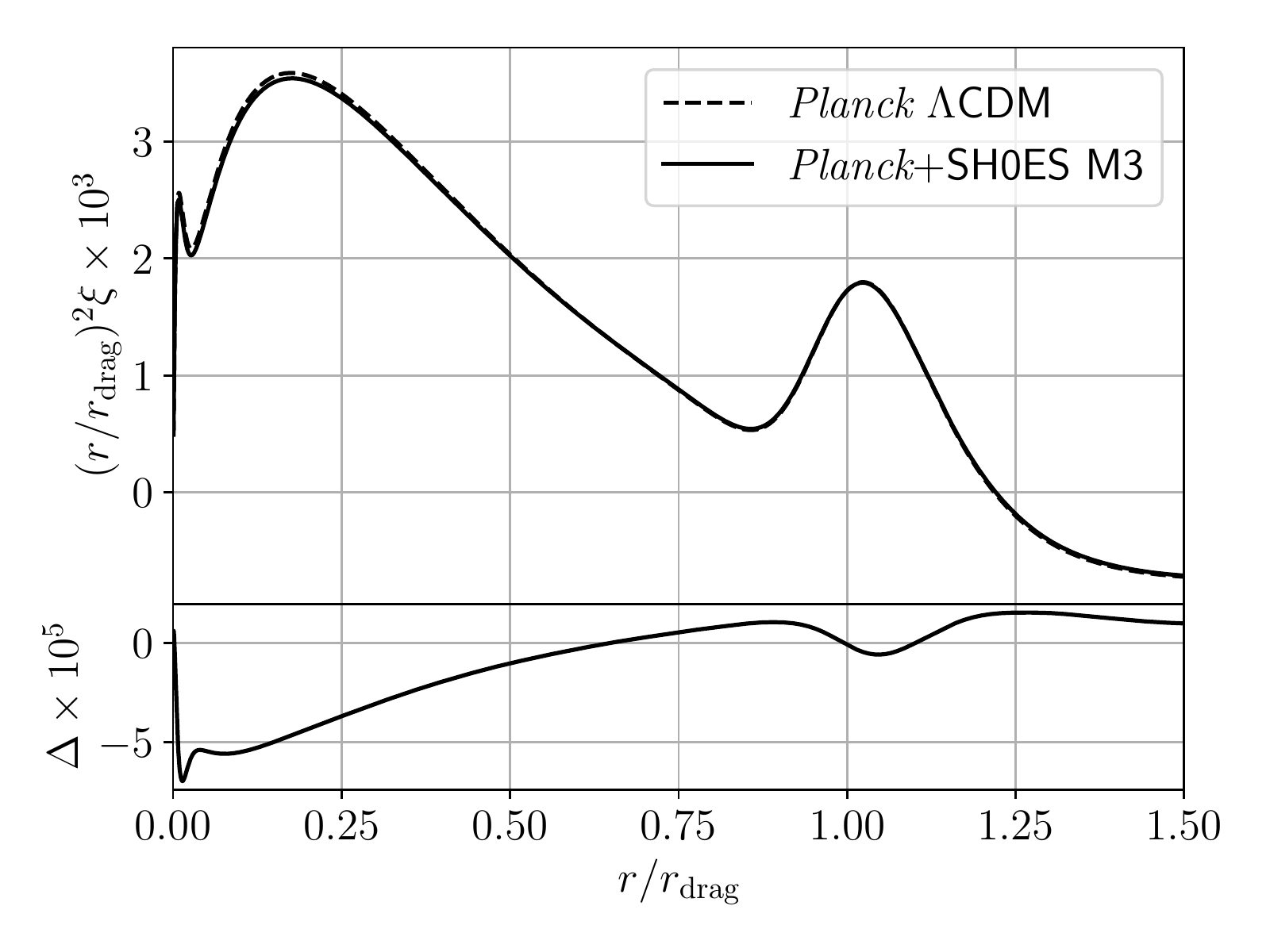}
\caption{Correlation functions at $z=0$ for $\Lambda$CDM (best fit to {\it Planck}) and for M3 (best fit to {\it Planck}+SH0ES), rescaled with each baryon drag scale, which gives almost perfect agreement.
The lower panel shows the difference between the two lines magnified by an additional factor of 100.
The comoving drag scale $r_{\rm drag}$ decreases by 1\% between the two cases, but $h$ increases by 4\%, so that the drag scale in Mpc/$h$ units $hr_{\rm drag}$ increases by 3\%.
Such change is clearly noticeable and overwhelms any possible bias arising from template-shape differences.}
\label{fig:bestfits-corfunc}
\end{figure}

We note that the change in recombination induced by small-scale clumping is likely to affect the shape of BAO fitting templates, and henceforth the distance-scale extraction from observational data.
A proper analysis of BAO data within our M3 model should check whether $d_M\left(z\right)/r_{\rm drag}$ and $H\left(z\right)r_{\rm drag}$ are recovered without any biases compared to standard extraction procedures.
A quick test with $z=0$ correlation functions in Fig.~\ref{fig:bestfits-corfunc} shows that change in $hr_{\rm drag}$ overwhelms the possible bias in drag scale reconstruction.
We leave a detailed study of the correlation function in the presence of clumping for future work.

\section{Forecasts for future CMB experiments}
\label{sec:forecasts}

Having exploited the current data at our disposal, we now perform forecasts for two future CMB experiments: the {\it Simons Observatory} (SO, \cite{SO}) and CMB-S4 \citep{CMBS4}, which will have much better damping-tail precision and therefore will be able to test the clumping model much better.

We have written mock likelihoods for these two experiments within Cobaya\footnote{\url{https://github.com/misharash/cobaya\_mock\_cmb}} adapting the ones in MontePython \citep{montepython1,montepython3}, and created the models for these two experiments using deproj0 noise curves for temperature and E-mode polarization fluctuations. 
We focus on primary CMB anisotropies, rather than lensing map.

\begin{table}[ht!]
\centering
\begin{tabular}{|c|c|c|}
\hline
 & Standard & Clumping \\
\hline
best fit to & {\it Planck} & +SH0ES \\
\hline
$\delta_-$ & n/a (0) & $-$0.955 \\
$\delta_+$ & n/a (0) & 1.320 \\
$f_0$ & n/a (1) & 0.652 \\
\hline
$b$ & n/a (0) & 0.439 \\
\hline
$10^9 A_s$ & 2.1094 & 2.1132 \\
$n_s$ & 0.96604 & 0.96552 \\
$100\theta_s$ & 1.04192 & 1.04177 \\
$\Omega_b h^2$ & 0.022416 & 0.022714 \\
$\Omega_\mathrm{cdm} h^2$ & 0.11945 & 0.11999 \\
$\tau_\mathrm{reio}$ & 0.0514 & 0.0542 \\
\hline
$H_0$ [km/(s Mpc)] & 68.146 & 70.916 \\
\hline
$\Omega_K$ & \multicolumn{2}{c|}{0} \\
$m_\nu$ [eV] & \multicolumn{2}{c|}{0} \\
\hline
\end{tabular}
\caption{Fiducial parameters for our forecasts.}
\label{tab:fiducials}
\end{table}

Throughout this section we consider two fiducial power spectra, which bracket our current knowledge on clumping during recombination: for the first we assume standard recombination ($\Lambda$CDM) and take CMB + direct $H_0$ measurement, whereas for the second we choose a model with nonzero clumping and higher $H_0$ and consider only CMB.
Full parameter sets are presented in Table \ref{tab:fiducials}.
For each we perform model comparison between M3 and LCDM, and show posteriors for $H_0$ and the clumping parameter $b$.
For the $H_0$ measurement, we assume SH0ES, though we have checked that an $H_0$ precision improvement to 1\% will not change our conclusions.

\subsection{Fiducial with standard recombination}
\label{sec:forecasts-standard}

We begin by considering the case that the CMB power spectra of SO/CMB-S4 continue to agree with the standard $\Lambda$CDM model [and a low $H_0\approx 68.1$ km/(s Mpc), where the parameters are taken from our best fit to {\it Planck} data with massless neutrinos and presented in Table~\ref{tab:fiducials}].
The question we address is whether the degeneracy between clumping and $H_0$ will still be able to bring future CMB experiments in closer agreement with a direct $H_0$ measurement in this case.
Since our fiducial is $\Lambda$CDM with standard recombination, M3 can not fit the data any better, so model comparison on CMB-only data will not be informative.
Therefore, in this subsection we consider only future CMB data added to SH0ES.

We show the model comparison between M3 and $\Lambda$CDM in Table \ref{tab:model-comparison-forecast-standard}.
Unlike with {\it Planck}, SO/CMB-S4+SH0ES have negligible $\chi^2$ improvement.
Bayes factors $K$ stay consistent with 1, indicating no clear preference between the models.

\begin{table}[ht!]
\centering
\begin{tabular}{|c|c|c|}
\hline
 & $\Delta \chi^2_{best}$ & $\log_{10} K$ \\
\hline
{\it Planck}+SH0ES & 5 & $0.14\pm 0.19$ \\
\hline
SO baseline+SH0ES & 0 & $0.15\pm 0.22$ \\
CMB-S4+SH0ES & 0 & $-0.02\pm 0.22$ \\
\hline
\end{tabular}
\caption{Model comparison forecast with a standard recombination fiducial (see Sec.~\ref{sec:forecasts-standard}).
All $\chi^2$ differences are rounded to integers because of uncertainty in the minimizer output.
If CMB data continues to be consistent with standard recombination and low $H_0$, M3 will be not able to allow CMB to agree with a direct $H_0$ measurement.
Bayes factors $-0.5\lesssim\log_{10}K\lesssim 0.5$ tell no clear preference, as in Tables \ref{tab:model-comparison-current} and \ref{tab:model-comparison-lowl}.}
\label{tab:model-comparison-forecast-standard}
\end{table}

\begin{figure}[ht!]
\includegraphics[width=\columnwidth]{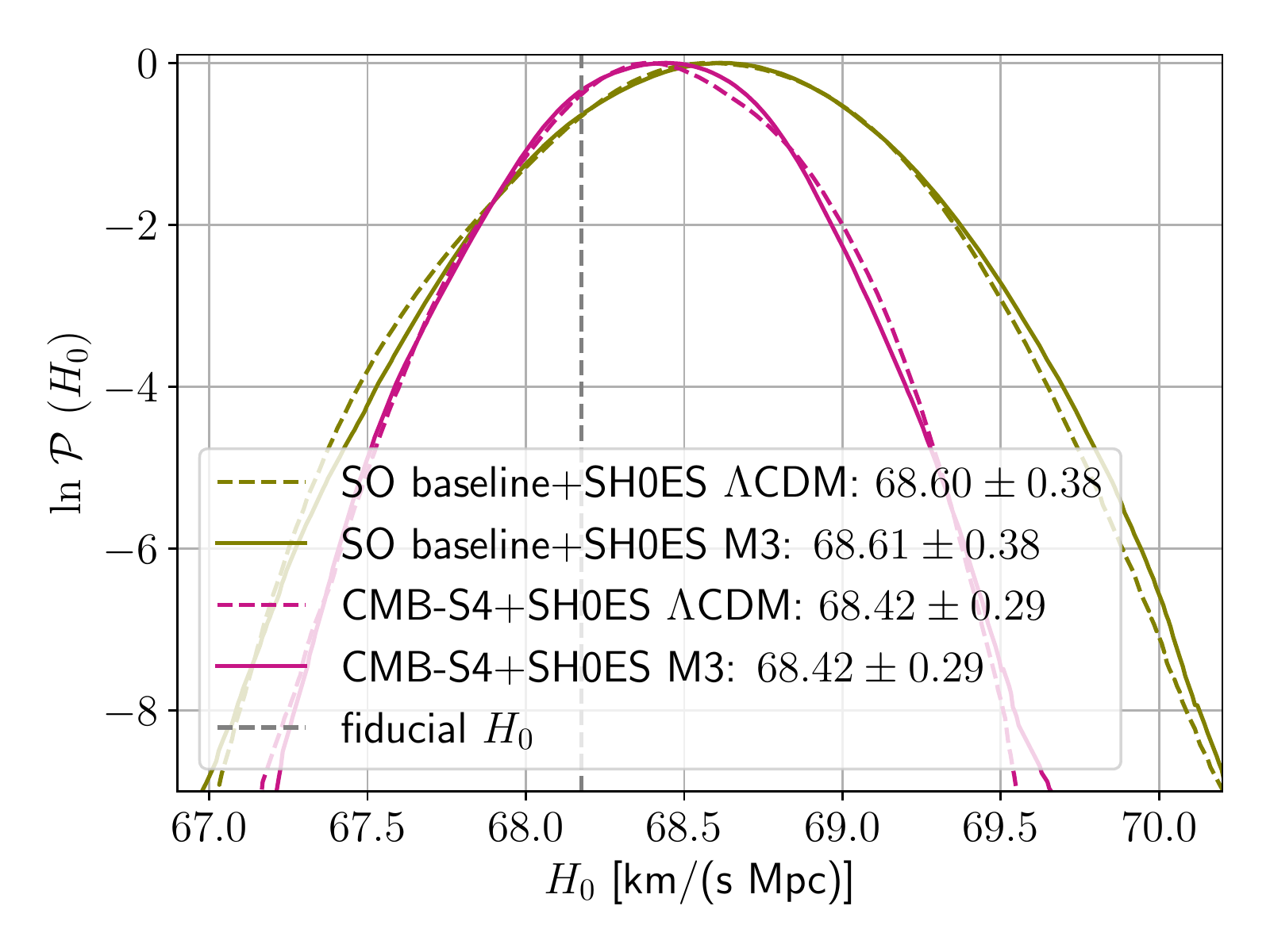}
\caption{Posterior for $H_0$ in our forecasts with a standard recombination fiducial, in all cases with CMB + SH0ES.
The solid lines assume M3, whereas dashed assume $\Lambda$CDM with standard recombination.
The gray dashed line shows the input fiducial value of $H_0$.
While the maximum posterior values are shifted to higher $H_0$ with respect to the input, due to the addition of a direct $H_0$ measurement, the shift from $\Lambda$CDM to M3 is negligible for both experiments.
We present 68\% C.L. intervals on $H_0$ in the legend.
For clumping (within M3) they are $\log_{10}\,b = -5.3^{+4.2}_{-3.7}$ for SO+SH0ES, $\log_{10}\,b = -5.8^{+3.7}_{-3.2}$ for CMB-S4+SH0ES (and $\log_{10}\,b = -5.5\pm 3.0$ for prior).}
\label{fig:H0-CMBSH0ES}
\end{figure}

Figure \ref{fig:H0-CMBSH0ES} provides a closer look into the $H_0$ posterior.
If future CMB power spectra continue to agree with $\Lambda$CDM, M3 does not allow any significant $H_0$ shift even with the pull from SH0ES.
As expected, the increased CMB precision shifts the $H_0$ posterior to our input CMB fiducial of $H_0\approx 68$ km/(s Mpc), shown as the gray dashed vertical line.
The posterior for $b$ is very close to its prior, except that high values $b\sim 1$ are disfavored (and thus does not allow any correlations with $H_0$), so we do not show it here.

\subsection{Fiducial with clumping}
\label{sec:forecasts-clumping}

We next investigate a case that in truth contains substantial small-scale clustering.
We generate fiducial power spectra for SO/CMB-S4 following the best fit of M3 to {\it Planck}+SH0ES, which in particular has $b\approx 0.44$ and $H_0\approx 70.9$ km/(s Mpc) (full parameter set in Table \ref{tab:fiducials}).
In this case our aim is to determine how clearly clumping could be discerned by future CMB data, without any direct $H_0$ measurements.

\begin{table}[ht!]
\centering
\begin{tabular}{|c|c|c|c|}
\hline
 & $\Delta \chi^2_{\rm best}$ & $\log_{10} K$ \\
\hline
{\it Planck} & 0 & $0.04\pm 0.15$ \\
\hline
SO baseline & {\bf 21} & ${\bf 1.57\pm 0.23}$ \\
CMB-S4 & {\bf 44} & ${\bf 5.54\pm 0.23}$ \\
\hline
\end{tabular}
\caption{Same as Table \ref{tab:model-comparison-forecast-standard}, but for a nonzero clumping fiducial (see Sec.~\ref{sec:forecasts-clumping}). In this case both experiments can show a clear preference for M3, given their large $\Delta \chi^2_{\rm best}$.
In terms of the Bayes factor $K$, the evidence against $\Lambda$CDM with standard recombination would be strong for SO ($1<\log_{10}K<2$), and decisive for CMB-S4 ($\log_{10}K>2$) \citep{bayes-factors}.
}
\label{tab:model-comparison-forecast-clumping}
\end{table}

We present the results of our model comparison in Table \ref{tab:model-comparison-forecast-clumping}.
If there is such clumping in CMB, SO data will show a clear preference for clumping model, and CMB-S4 will be even more decisive.
This is the only case when the Bayes factor $K$ is significantly different from 1.
With SO data M3 model is deemed $\sim 30$ times more probable than $\Lambda$CDM (strong evidence in favor of M3), and with CMB-S4---$\sim 300,000$ (decisive evidence in favor of M3) \citep{bayes-factors}.

\begin{figure}[ht!]
\includegraphics[width=\columnwidth]{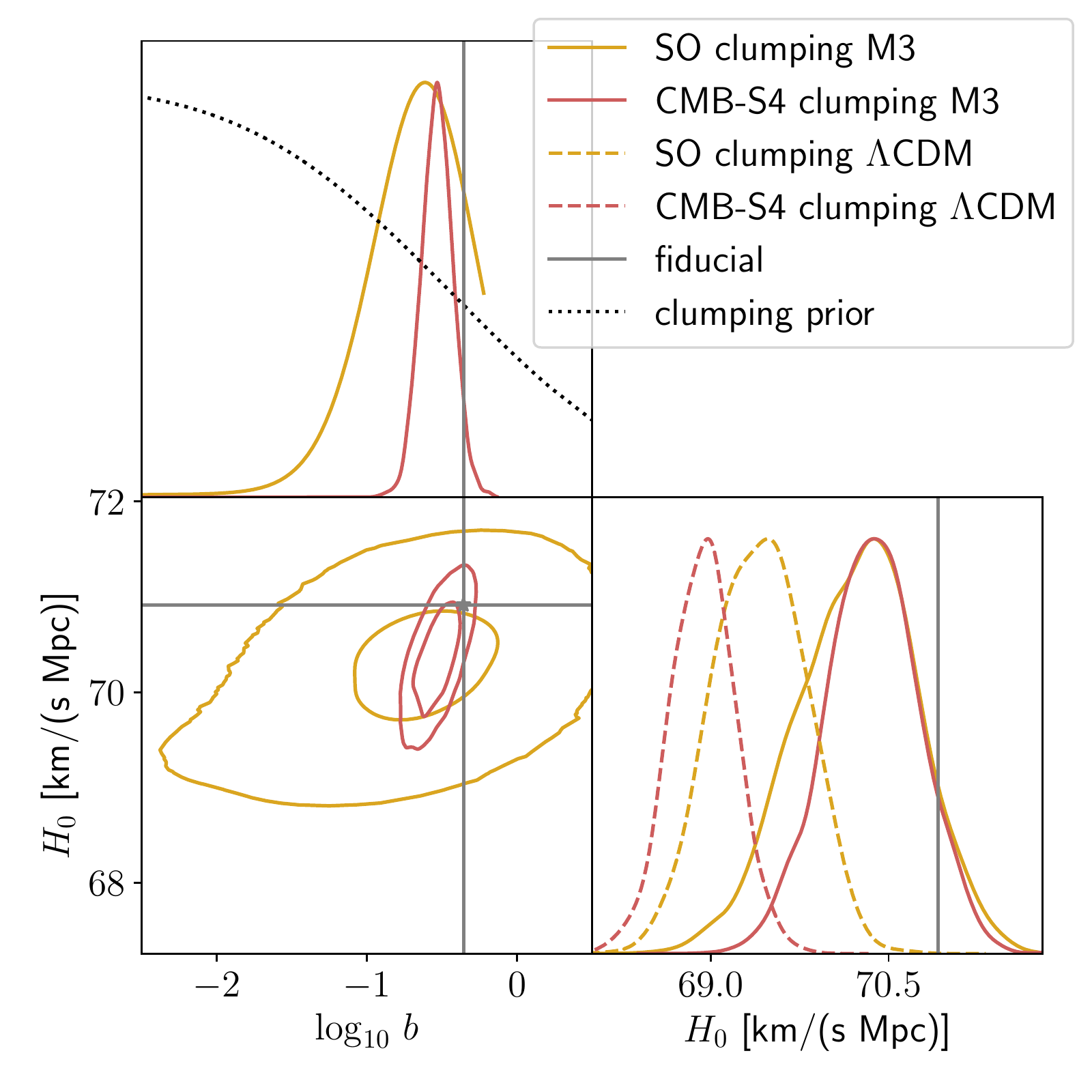}
\caption{Posteriors of $b$ and $H_0$ for forecasts with a clumping fiducial (best fit to current {\it Planck}+SH0ES, see Sec.~\ref{sec:forecasts-clumping}).
As is clear from the right panel, M3 and $\Lambda$CDM yield different $H_0$ values for the same data (both for SO and CMB-S4).
In this case the clumping parameter $b$ will be measured to be nonzero at high significance.
The 68\% C.L. intervals on clumping within M3 are $\log_{10}\,b = -0.93^{+0.51}_{+0.16}$ for SO, $\log_{10}\,b = -0.53\pm 0.11$ for CMB-S4 (and $\log_{10}\,b = -5.5\pm 3.0$ for prior).
Within M3, $H_0 = 70.23^{+0.54}_{-0.47}$ km/(s Mpc) for SO and $H_0 = (70.36\pm 0.39)$ km/(s Mpc) for CMB-S4.
}
\label{fig:forecasts-clumping-b-H0}
\end{figure}

We show the posteriors of $b$ and $H_0$ in Fig.~\ref{fig:forecasts-clumping-b-H0}, where the difference in $H_0$ between standard and clumpy recombination is clear for both SO and CMB-S4.
We note that the posteriors for both $H_0$ and $b$ peak at lower values than our input fiducials, as lower clumping (and therefore lower $H_0$ for the same $\theta_s$) is favored by the prior.
Also note that all results in this subsection are based on CMB data only, without any direct $H_0$ measurements.
It is clear that future CMB data alone will suffice to detect clumping, as the posterior becomes much better constrained (against the prior).
For SO the lowest values ($b\lesssim 10^{-2}$) are clearly disfavored by the data, whereas for CMB-S4 the limits become only tighter.

\section{Conclusions}
\label{sec:conclusion}

The Hubble tension poses an increasingly challenging problem to the standard cosmological model.
A possible solution is to alter recombination, for instance by adding
small-scale baryon clumping, which allows higher $H_0$ values to be inferred from CMB data.
We have studied whether our flexible clumping model M3, having three spatial zones with variable densities and volume fractions, can solve the tension.

We have found that
\renewcommand{\theenumi}{(\roman{enumi})}
\renewcommand{\labelenumi}{\theenumi}
\begin{enumerate}
\item Current {\it Planck} data does not prefer clumping, even when adding the local $H_0$ measurement from the SH0ES Collaboration.
\item Including only $\ell<1000$ multipoles, {\it Planck} data allow for a larger shift to higher values of $H_0$, as the damping tail is more weakly constrained.
\item Increasing $H_0$ within $\Lambda$CDM decreases both $\Omega_m$ and $S_8$. The clumping model M3 follows the same trend, which relieves the potential $S_8$ tension with weak-lensing data.
\item However, the same change of $\Omega_m$ is in tension with BAO standard-ruler measurements at low $z$. 
We showed that the BAO template is largely unaltered in the presence of clumping.
\end{enumerate}

We have made forecasts for two future CMB experiments---{\it Simons Observatory} and CMB-S4---which will better measure the damping tail.
First, we have found that if the power spectra stay consistent with $\Lambda$CDM (i.e., with standard recombination), increasing $H_0$ via clumping is strongly disfavored.
Second, we have shown that the current best-fit model to {\it Planck}+SH0ES with clumping ($b\approx 0.4$) can be detected at high significance based solely on future CMB data.
Therefore future CMB experiments will provide considerable diagnostic power to investigate small-scale clumping at the epoch of recombination and shed light onto possible solutions to the $H_0$ tension.
	
\section{Acknowledgements}
JBM was funded through a Clay fellowship at the Smithsonian Astrophysical Observatory.
DJE is partially supported by U.S. Department of Energy Grant No. DE-SC0013718 and as a Simons Foundation Investigator.
CD is partially supported by the Department of Energy (DOE) Grant No. DE-SC0020223.

\bibliographystyle{apsrev4-2_16.bst}
\bibliography{references.bib}

\appendix

\section{JUSTIFYING SIMPLIFICATIONS}

\subsection{RECFAST vs HyREC}
\label{sec:justify-reccodes}

\begin{figure}[ht!]
\includegraphics[width=\columnwidth]{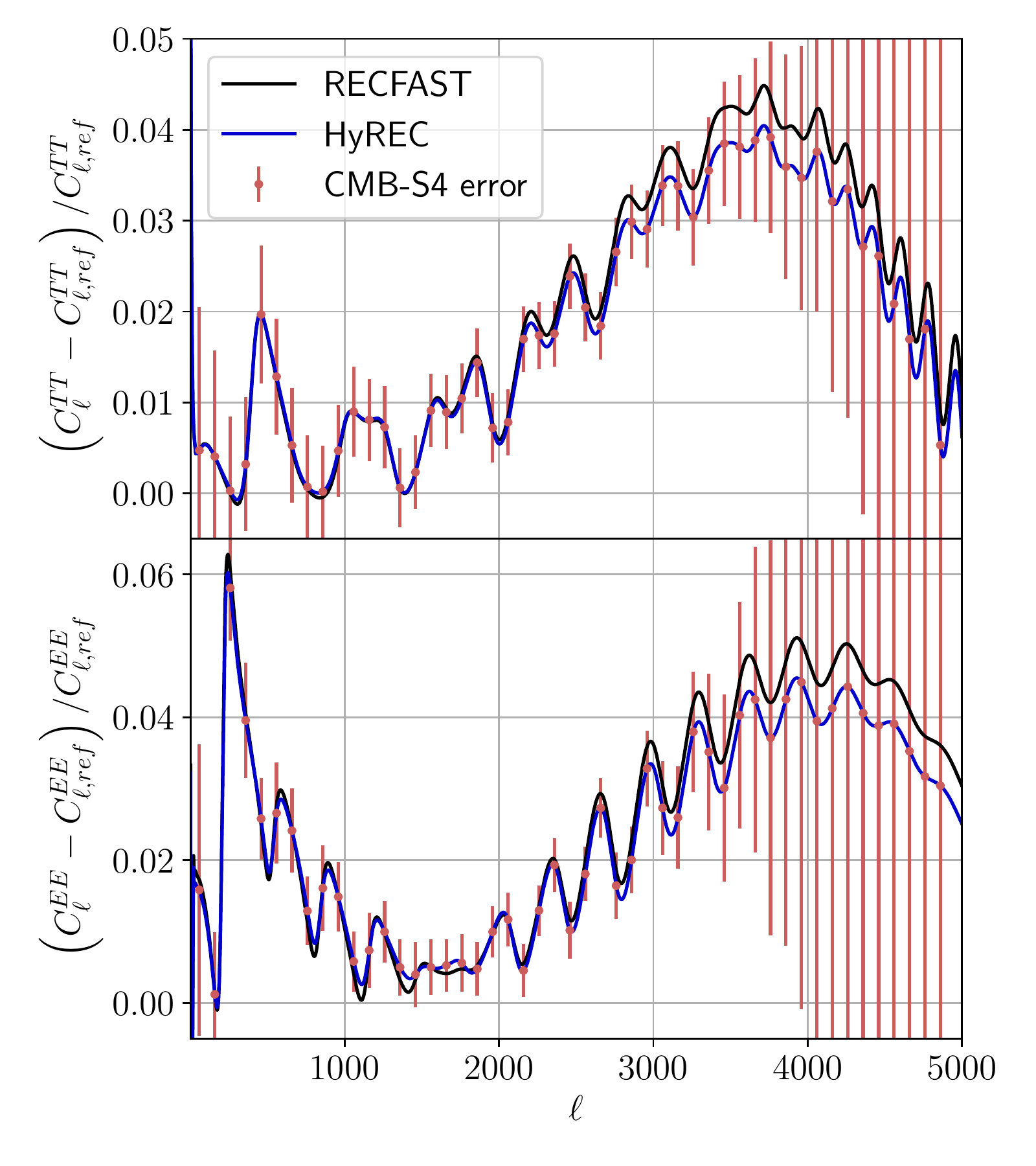}
\caption{Relative shifts between $\Lambda$CDM with standard recombination (taken as reference $C_{\ell,\rm ref}$) and a particular M3 configuration ($\delta_-=-0.9$, $\delta_+=5/3$, $f_0=1/3$ giving $b=1$, like in Fig.~\ref{fig:M3demo-damping}), using RECFAST (in black) and HyREC (in blue).
The red bands correspond to CMB-S4 errors binned with $\Delta\ell=100$.
The cosmological parameters ($\theta_s$, $\omega_b$, $\omega_\mathrm{cdm}$, $A_s$, $n_s$, $\tau_\mathrm{reio}$) are fixed to the Planck best fit. }
\label{fig:M3demo-reccodes}
\end{figure}

In the main text we used the recombination code RECFAST, as it is faster than the more precise HyREC; here we justify that this choice does not bias our results.
RECFAST is sufficiently accurate for the analysis of {\it Planck} data, though this code will not be satisfactory for future CMB missions \citep{hyrec2}.
Moreover, highly nonstandard hydrogen densities, which can appear in the $\pm$ zones in M3, might limit the RECFAST applicability even further.

For our purposes, the change of zero-point is not the most important, as we focus on the shift introduced by clumping.
Therefore, we compare the relative changes in $C_\ell^{TT/EE}$ between $\Lambda$CDM with standard recombination and M3 with $b=1$ clumping obtained with RECFAST and HyREC (with full hydrogen model) in Fig.~\ref{fig:M3demo-reccodes}.
We overlay the CMB-S4 error bars to assess the difference.
There is no notable difference for $\ell\leq 2000$-2500, so for current {\it Planck} data both codes can be considered equivalent.
In particular, the differences between each M3 model and $\Lambda$CDM are $\Delta\chi^2_{Planck,\rm RECFAST}=93$ and $\Delta\chi^2_{Planck,\rm HyREC}=90$, which are very close (as we note we have not shifted any parameters here).
For SO or CMB-S4, however, the difference between the recombination codes can be larger. 
For the lines shown, $\Delta\chi^2_{\rm CMB-S4,RECFAST}=1350$ and $\Delta\chi^2_{\rm CMB-S4,HyREC}=1150$, if one assumes the {\it Planck} best-fit cosmology (fixed), which shows a relative difference between the recombination codes of $\lesssim 20\%$.
Near the fiducial, where most of posterior is, the absolute difference will naturally be lower.
Also, for CMB-S4 precision the shifts in parameters are expected to be less than $1\sigma$ \citep{hyrec2}.
We conclude that an analysis of real SO and CMBS4 data  should use HyREC, though RECFAST is sufficient for our forecasting purposes.

\subsection{Neutrino masses}
\label{sec:justify-mnu}

\begin{figure}[ht!]
\includegraphics[width=\columnwidth]{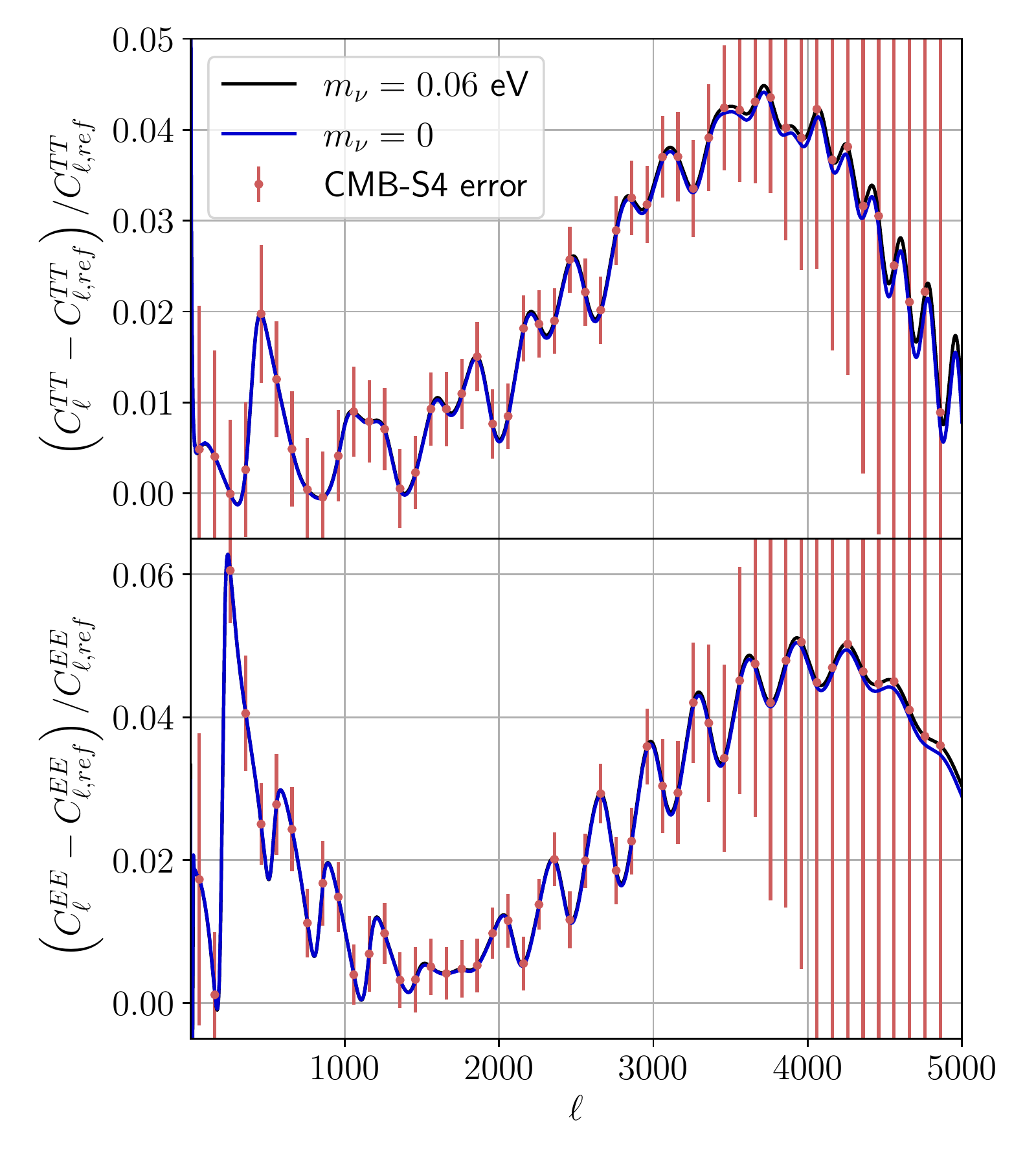}
\caption{Same as Fig.~\ref{fig:M3demo-reccodes} but comparing the cases of massive (black) and massless (blue) neutrinos.}
\label{fig:M3demo-neutrinos}
\end{figure}

Throughout this paper we assumed massless neutrinos for efficiency, as it reduces the computational overhead by an order of magnitude. 
This increases our best fit $H_0$ compared to the {\it Planck} one.
However, again, we are most interested in changes introduced by clumping with respect to standard recombination, so we compare them for massive and massless neutrinos in Fig.~\ref{fig:M3demo-neutrinos}.
The difference between both predictions in this plot is minuscule, and always smaller than even the CMB-S4 error bars.
More quantitetively, for the lines shown, $\Delta\chi^2_{Planck,m_\nu=0.06\,\rm eV}=92.6$ and $\Delta\chi^2_{Planck,m_\nu=0\,\rm eV}=93.3$; $\Delta\chi^2_{{\rm CMB-S4},m_\nu=0.06\,\rm eV}=1346$ and $\Delta\chi^2_{{\rm CMB-S4},m_\nu=0\,\rm eV}=1334$ (if one assumes {\it Planck} best fit cosmology for fiducial).
The relative difference in $\Delta\chi^2$ is $\lesssim 1\%$ in both cases.
Therefore computing with massless neutrinos suffices our purposes.

\section{FULL CONTOURS FROM {\it PLANCK} RUNS}
\label{sec:fullcontours}

\begin{figure*}[ht!]
\includegraphics[width=\textwidth]{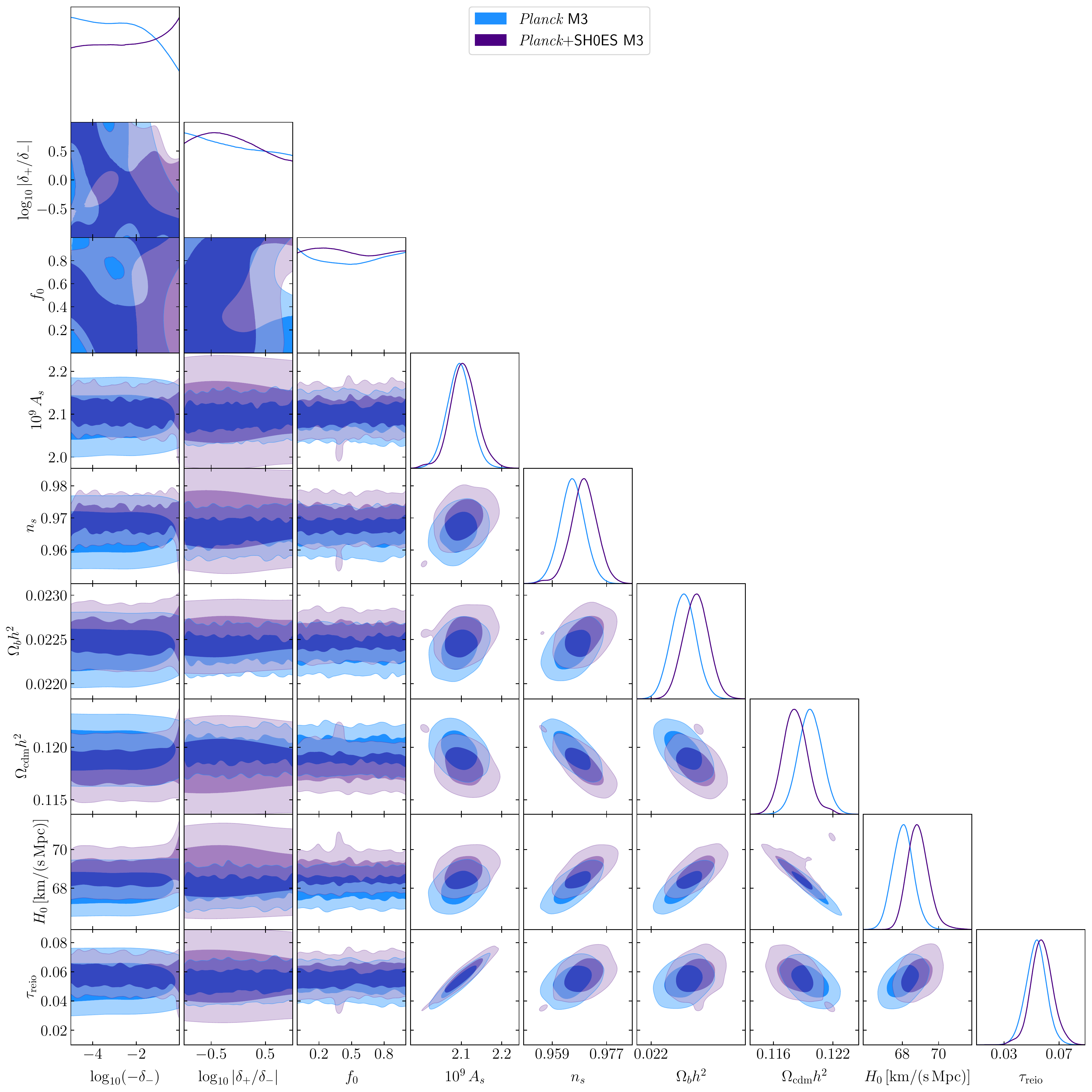}
\caption{Full contour plot for {\it Planck} runs of M3. Purple is with SH0ES, blue is without.}
\label{fig:planckSH0ES-fullcontours}
\end{figure*}

In Fig.~\ref{fig:planckSH0ES-fullcontours} we show posteriors for all parameters in runs of M3 with {\it Planck} 2018 data (without and with SH0ES).
The posteriors on clumping parameters $\log_{10}\left(-\delta_-\right)$, $\log_{10}\left|\delta_+/\delta_-\right|$ and $f_0$ are largely flat, like the priors, except the decrease for higher $|\delta_-|$ for {\it Planck} and increase in the same place for {\it Planck}+SH0ES.
The clumping parameters also show almost no correlations with standard cosmological parameters, except a weak $H_0$ increase for the highest $|\delta_-|$.
Addition of SH0ES causes some increase in $n_s$, $\Omega_bh^2$, $H_0$; a weak increase in $A_s$ and $\tau_{\rm reio}$; some decrease in $\Omega_{\rm cdm}h^2$.

\section{BEST FIT PARAMETERS}
\label{sec:bestfits}

\begin{table}[ht!]
\centering
\begin{tabular}{|c|c|c|c|}
\hline
Model & $\Lambda$CDM & $\Lambda$CDM & M3 \\
\hline
Fit to & {\it Planck} & \multicolumn{2}{c|}{{\it Planck}+SH0ES} \\
\hline
$\delta_-$ & n/a (0) & n/a (0) & $-$0.955 \\
$\delta_+$ & n/a (0) & n/a (0) & 1.320 \\
$f_0$ & n/a (1) & n/a (1) & 0.652 \\
\hline
$b$ & n/a (0) & n/a (0) & 0.439 \\
\hline
$10^9 A_s$ & 2.1094 & 2.1440 & 2.1132 \\
$n_s$ & 0.96604 & 0.97084 & 0.96552 \\
$100\theta_s$ & 1.04192 & 1.04207 & 1.04177 \\
$\Omega_b h^2$ & 0.022416 & 0.022569 & 0.022714 \\
$\Omega_\mathrm{cdm} h^2$ & 0.11945 & 0.11762 & 0.11999 \\
$\tau_\mathrm{reio}$ & 0.0514 & 0.0555 & 0.0542 \\
\hline
$H_0$ [km/(s Mpc)] & 68.146 & 68.993 & 70.916 \\
\hline
$\Omega_K$ & \multicolumn{3}{c|}{0} \\
$m_\nu$ [eV] & \multicolumn{3}{c|}{0} \\
\hline
$\chi^2_{{\rm low}\,\ell\,TT}$ & 23.2 & 22.4 & 23.6 \\
$\chi^2_{{\rm low}\,\ell\,EE}$ & 395.7 & 396.1 & 395.9 \\
$\chi^2_{{\rm high}\,\ell\,TTTEEE}$ & 582.2 & 584.0 & 585.4 \\
$\chi^2_{\rm lensing}$ & 9.0 & 8.7 & 8.8 \\
\hline
$\chi^2_{\it Planck}$ & 1010.0 & 1011.1 & 1013.7 \\
\hline
$\chi^2_{\rm SH0ES}$ & (15.1) & 10.5 & 3.1 \\
\hline
\end{tabular}
\caption{Our best fit parameters to (full) {\it Planck} and {\it Planck}+SH0ES.}
\label{tab:bestfits-planck}
\end{table}

In Table~\ref{tab:bestfits-planck} we show the best fit parameters when fitting with full {\it Planck} data.
The M3 best fit to {\it Planck}-only is not shown, as we have not found a better one than $\Lambda$CDM, and the $\Lambda$CDM is included in M3 when one sets either one of the $\delta$'s to 0 or $f_0=1$.
Adding SH0ES in $\Lambda$CDM naturally increases $H_0$ as well as most input parameters: $A_s$, $n_s$, $\theta_s$, $\omega_b$, $\tau_\mathrm{reio}$; whereas $\omega_\mathrm{cdm}$ decreases slightly.
M3 allows for larger $H_0$, while the increase in $A_s$ and $\tau_\mathrm{reio}$ become smaller, $n_s$ and $\theta_s$ decrease very slightly, $\Omega_bh^2$ increases further than in $\Lambda$CDM and $\Omega_{\rm cdm}h^2$ increases, unlike in $\Lambda$CDM.
The CMB $\chi^2$ difference is dominated by high-$\ell$ $TT,TE,EE$ (where ``high'' means $\ell>30$).

\begin{table}[ht!]
\centering
\begin{tabular}{|c|c|c|c|}
\hline
Model & $\Lambda$CDM & $\Lambda$CDM & M3 \\
\hline
Fit to & {\it Planck} $\ell<1000$ & \multicolumn{2}{c|}{{\it Planck} $\ell<1000$ + SH0ES} \\
\hline
$\delta_-$ & n/a (0) & n/a (0) & $-$0.950 \\
$\delta_+$ & n/a (0) & n/a (0) & 1.196 \\
$f_0$ & n/a (1) & n/a (1) & 0.301 \\
\hline
$b$ & n/a (0) & n/a (0) & 0.794 \\
\hline
$10^9 A_s$ & 2.0918 & 2.1315 & 2.0705 \\
$n_s$ & 0.97026 & 0.97711 & 0.95944 \\
$100\theta_s$ & 1.04150 & 1.04179 & 1.04034 \\
$\Omega_b h^2$ & 0.022537 & 0.022773 & 0.022700 \\
$\Omega_\mathrm{cdm} h^2$ & 0.11831 & 0.11613 & 0.12198 \\
$\tau_\mathrm{reio}$ & 0.0518 & 0.0553 & 0.0505 \\
\hline
$H_0$ [km/(s Mpc)] & 68.528 & 69.640 & 72.616 \\
\hline
$\Omega_K$ & \multicolumn{3}{c|}{0} \\
$m_\nu$ [eV] & \multicolumn{3}{c|}{0} \\
\hline
$\chi^2_{{\rm low}\,\ell\,TT}$ & 22.4 & 21.4 & 24.8 \\
$\chi^2_{{\rm low}\,\ell\,EE}$ & 395.7 & 395.9 & 395.6 \\
$\chi^2_{{\rm high}\,\ell\,TTTEEE}$ & 286.3 & 288.2 & 281.5 \\
$\chi^2_{\rm lensing}$ & 8.9 & 9.0 & 8.8 \\
\hline
$\chi^2_{\it Planck}$ & 713.2 & 714.4 & 710.7 \\
\hline
$\chi^2_{\rm SH0ES}$ & (12.9) & 7.5 & 0.2 \\
\hline
\end{tabular}
\caption{Our best fit parameters to {\it Planck} $\ell<1000$, without and with SH0ES.}
\label{tab:bestfits-planck-lowl}
\end{table}

In Table~\ref{tab:bestfits-planck-lowl} we show the best fit parameters when fitting with {\it Planck} $\ell<1000$ data.
The shifts in cosmological parameters are similar and small.
Interestingly, the addition of SH0ES helped to find a better fit to {\it Planck} than $\Lambda$CDM, unlike with {\it Planck}-only data.
This is likely because the optimal parameters region with high clumping is small.
However, the fit improvement is not significant.
The CMB $\chi^2$ difference is also dominated by high-$\ell$ $TT,TE,EE$ (where ``high'' now means $30<\ell<1000$).

\end{document}